\pdfoutput=1

\documentclass[a4paper,fleqn,usenatbib]{mnras}

\usepackage{mathptmx}

\usepackage[T1]{fontenc}
\usepackage{ae,aecompl}

\usepackage{graphicx}	
\usepackage{amsmath}	
\usepackage{amssymb}	
\usepackage{cite}
\usepackage{caption}
\usepackage{subfig}
\usepackage{epstopdf}

\newcommand{\msun}{M$_\odot$}
\newcommand{\spcu}{\,cts s$^{-1}$ PCU$^{-1}$}
\newcommand{\rxte}{\textit{RXTE}}
\newcommand{\pca}{\textit{PCA}}
\newcommand{\ergf}{\,ergs~s$^{-1}$~cm$^{-2}$}

\title[Exotic Variability in IGR J17091-3624]{An Atlas of Exotic Variability in IGR J17091-3624: A Comparison with GRS 1915+105}

\author[J.M.C. Court et al.]{
J.M.C. Court$^{1}$\thanks{E-mail: J.M.Court@soton.ac.uk},
D. Altamirano$^{1}$,
M. Pereyra$^{1}$,
C.M. Boon$^{1}$,
K. Yamaoka$^{2}$,
\newauthor T. Belloni$^{3}$,
R. Wijnands$^{4}$,
M. Pahari$^{5}$
\\
$^{1}$Department of Physics and Astronomy, University of Southampton, Southampton, SO17 1BJ, UK\\
$^{2}$Department of Physics, Nagoya University, Aichi 464-8602, Japan\\
$^{3}$Osservatorio Astronomico di Brera, Via E. Bianchi 46, 23807 Merate (LC), Italy\\
$^{4}$Anton Pannekoek Institute for Astronomy, University of Amsterdam, Postbus 94249, 1090 GE Amsterdam, The Netherlands\\
$^{5}$Inter-University Center for Astronomy and Astrophysics, Post Bag 4, Ganeshkhind, Pune-411007, India\\
}

\date{Accepted 2017 March 27. Received 2017 February 28; in original form 2016 November 25}

\pubyear{2017}

\begin{document}
\label{firstpage}
\pagerange{\pageref{firstpage}--\pageref{lastpage}}
\maketitle

\begin{abstract}
We performed an analysis of all \rxte\ observations of the Low Mass X-ray Binary and Black Hole Candidate IGR J17091-3624 during the 2011-2013 outburst of the source.  By creating lightcurves, hardness-intensity diagrams and power density spectra of each observation, we have created a set of 9 variability `classes' that phenomenologically describe the range of types of variability seen in this object.  We compare our set of variability classes to those established by \citet{Belloni_GRS_MI} to describe the similar behaviour of the LMXB GRS 1915+105, finding that some types of variability seen in IGR J17091-3624 are not represented in data of GRS 1915+105.  We also use all available X-ray data of the 2011-2013 outburst of IGR J17091-3624 to analyse its long-term evolution, presenting the first detection of IGR J17091-3624 above 150\,keV as well as noting the presence of `re-flares' during the latter stages of the outburst.  Using our results we place new constraints on the mass and distance of the object, and find that it accretes at $\lesssim33$\% of its Eddington limit.  As such, we conclude that Eddington-limited accretion can no longer be considered a sufficient or necessary criterion for GRS 1915+105-like variability to occur in Low Mass X-Ray Binaries.
\end{abstract}




\section{Introduction}

\par X-ray binaries are systems in which a black hole or neutron star accretes matter from a stellar companion, and they provide us with opportunities to test how accretion takes place in the most extreme physical regimes.  Some X-ray Binaries are believed to be accreting at very close to the Eddington limit, the limit at which the radiation pressure on accreting material is equal to the force due to gravity.  As such, these objects can also provide a laboratory with which to explore accretion in radiation pressure-dominated systems \citep{White_Radiation}.
\par Low Mass X-ray Binaries (hereafter LMXBs) are a subclass of X-ray binary in which the compact object accretes matter transferred to it due to a Roche-lobe overflow from the companion star (e.g. \citealp{Paczynski_Roche}).  In general, accretion in LMXBs is a variable process, with variability seen on timescales from milliseconds to decades.  On the shortest timescales the X-ray lightcurves of these objects can show band-limited noise and low-frequency quasi-periodic oscillations (QPOs) at frequencies from $\sim$mHz to $\sim$200\,Hz (e.g. \citealp{vanderKlis_QPOs}).  Black hole binaries also show so-called `high frequency QPOs' (e.g. \citealp{Remillard_HHz1,Remillard_HHz2,Belloni_QPOs,Belloni_Review}),  thought to be caused by motion of matter in the innermost region of the accretion disk (e.g. \citealp{Stefanov_HFQPOs}).
\par Three sources -- GRS 1915+105, IGR J17091-3624 and the neutron star `Rapid Burster' (MXB 1730-335) -- also show a variety of exotic variability on timescales of seconds to minutes in addition to the kinds of variability seen in other LMXBs.  This exotic variability consists of quasi-periodic flares, dips and other high-amplitude behaviours (e.g \citealp{Belloni_GRS_MI,Altamirano_IGR_FH,Bagnoli_RB}).  The second-to-minute-scale lightcurve profiles of these sources change over timescales of days.  In GRS 1915+105 and IGR J17091-3624, this behaviour can be described as a set of `variability classes'.  These classes themselves vary widely in terms of flux, structure, periodicity and spectral properties.
\par  GRS 1915+105 \citep{CastroTirado_GRS1915}, hereafter GRS 1915, is a black hole LMXB which accretes at between a few tens and 100\% of its Eddington Limit (e.g. \citealp{Vilhu_SupEd,Done_GRS_HighAcc,Fender_DiskJet,Reid_Parallax}).  Most LMXBs go through periods of low-intensity `quiescence' and high-intensity `outbursts', the latter consisting of black-body dominated `soft' and power-law dominated `hard' spectral states.  However, GRS 1915 has been in outburst since its discovery in 1992 \citep{CastroTirado_GRS1915}.  GRS 1915 is also notable for the incredible variety and complexity of variability classes it exhibits (e.g. \citealp{Yadav_GRSBursts,Belloni_GRS_MI}) in addition to the less exotic variability seen in other black hole binary systems.  GRS 1915 additionally shows high-frequency and low frequency QPOs similar to those seen in other black hole LMXBs \citep{Morgan_QPO}.  In total, 15 distinct variability classes have been observed \citep{Belloni_GRS_MI,KleinWolt_OmegaClass,Hannikainen_NewClass, Pahari_NewClass}.  This remarkable range of behaviour is believed to be caused by instability in the inner accretion disc (e.g. \citealp{Janiuk_RadInstab,Nayakshin_GRSModel}), which is in turn caused by the existence of a radiation pressure dominated regime in the inner disc (e.g. \citealp{Done_GRS_HighAcc}).   Accounting for this complexity could be key to our understanding of radiation-dominated accretion regimes.
\par One of the best-studied variability classes of GRS 1915 is the highly regular flaring $\rho$, or `heartbeat', class, so named for the similarity of its lightcurve to an electrocardiogram. It has been shown that hard X-ray photons tend to lag soft ones in this class (e.g. \citealp{Janiuk_Lag,Massaro_Lag}).  Numerical models derived from \citealp{Shakura_Disk} which reproduce this lag can also reproduce other flaring classes seen in GRS 1915 (e.g \citealp{Nayakshin_GRSModel,Massaro_Numerical}).  These numerical models predict that GRS 1915-like variability should be seen in systems accreting with a global Eddington fraction of $\gtrsim0.26$ \citep{Nayakshin_GRSModel}.  However, other LMXBs (e.g. GX 17+2, \citealp{Kuulkers_GX17}, and V404 Cyg, \citealp{Huppenkothen_V404}) have been observed to exceed this Eddington fraction without displaying GRS 1915-like variability.
\par
\begin{figure}
  \centering
  \includegraphics[width=.9\linewidth, trim= 25mm 0mm 0mm 0mm]{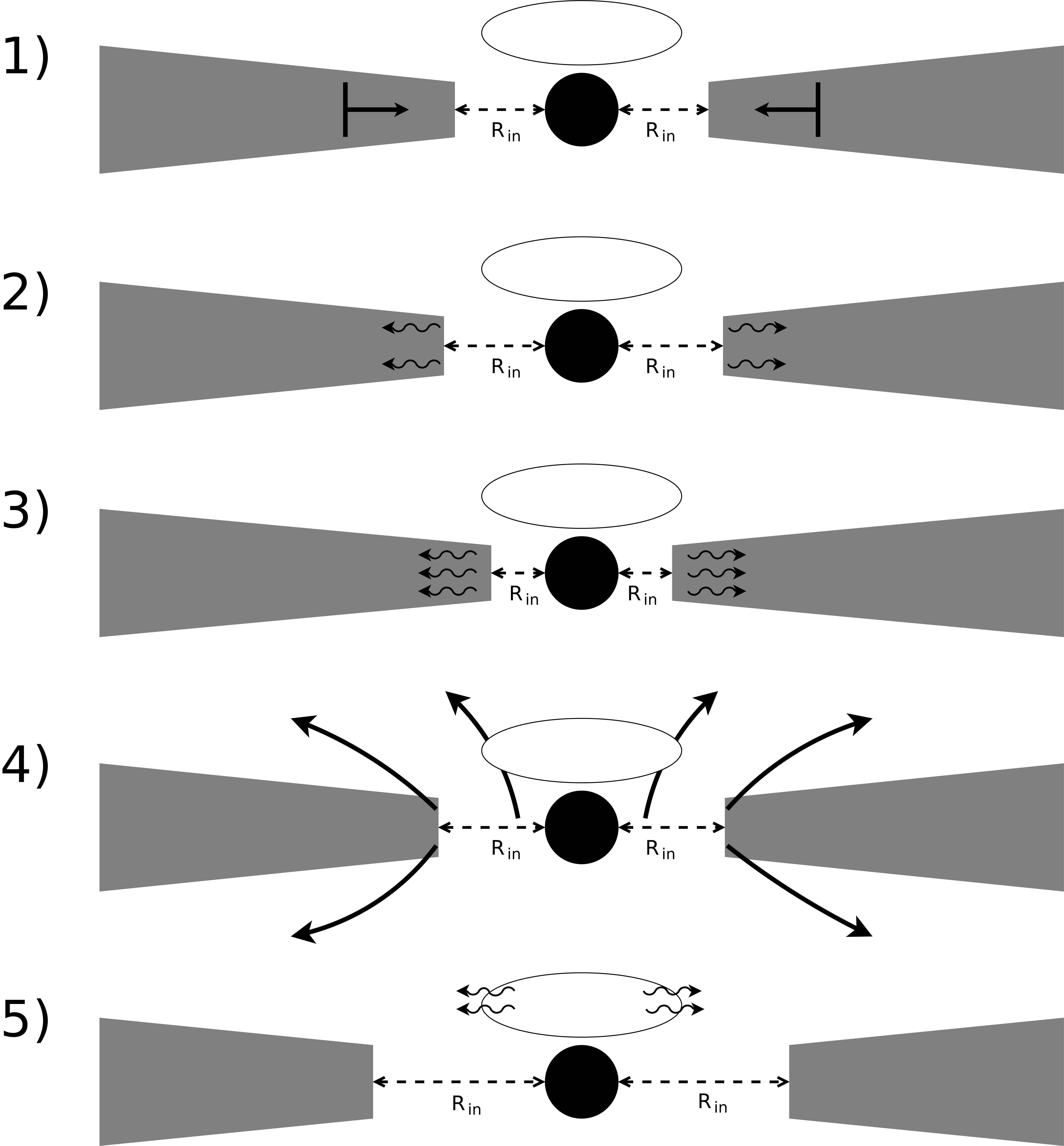}
  \caption{\small A schematic diagram illustrating the the process described by \citet{Neilsen_GRSModel} to describe the $\rho$ variability class in GRS 1915+105.  1) The X-ray emission from the system originates from both the accretion disc truncated at an inner radius $r_{in}$ (grey) and a cloud of non-thermal electrons (white ellipse).  At some time $t$, an overdensity in the accretion disc (formed by the Lightman-Eardley Instability) propagates inwards towards $r_{in}$.  2) As the inner disc heats up, $r_{in}$ begins to slowly increase due to an increase in photon pressure.  This destabilises the disc.  3) At some critical density, the disc becomes too unstable and collapses inwards, greatly decreasing $r_{in}$ and raising the inner disc temperature.  4) The sudden increase in emission exceeds the local Eddington limit at $r_{in}$, ejecting matter from the inner accretion disc in the form of extreme winds.  5) Having been excited by matter in the winds passing through it, the non-thermal electron cloud emits a hard Brehmsstrahlung `pulse'.}
  \label{fig:WindsModel}
\end{figure}

 \citet{Neilsen_GRSModel} proposed a physical scenario, based on the mathematical model proposed by \citet{Nayakshin_GRSModel}, to explain the presence of the hard lag in the flaring classes of GRS 1915.  This is outlined schematically in Figure \ref{fig:WindsModel}.  First, an overdensity of matter forms via the thermal-viscous Lightman-Eardley Instability \citep{Lightman_Instability} and propagates inwards through the accretion disc.  This destabilises the disc, collapsing its inner radius and vastly increasing photon emission.  If the local Eddington Limit in the inner accretion disc is then approached, extreme outflows are triggered that deplete the inner accretion disc and allow the cycle to begin again.  As the matter ejected from the disc collides with the non-thermal `corona' above the central object, a flash of hard Bremsstrahlung radiation is produced.  This causes a hardening of the spectrum and an apparent lag between soft and hard photons.  \citet{Janiuk_Lag} instead propose that the lag is caused by the corona smoothly adjusting to the changing brightness of the disc after a light travel time.

\par The black hole candidate LMXB IGR J17091-3624 (hereafter J17091) was discovered in outburst by \textit{INTEGRAL} in 2003 \citep{Kuulkers_IGRDiscovery}.  In 2011, it again went into outburst \citep{Krimm_IGROutburst}.  GRS 1915-like variability was discovered in its lightcurve, as well as high-frequency QPOs which behave much like the QPOs seen in GRS 1915 \citep{Altamirano_IGR_FH,Altamirano_Discovery,Altamirano_HFQPO}.  As IGR J17091 is around a factor of 20 fainter at 2--25\,keV than GRS 1915, this object has either a lower black hole mass $M$, a lower accretion rate $\dot{m}$ or lies at a larger distance $D$ than GRS 1915.  Assuming by analogy with GRS 1915 that IGR J17091 is accreting at its Eddington rate, the black hole must have a mass of $M\lesssim3$\msun\ or lie at a distance of $D\gtrsim20$ kpc \citep{Altamirano_IGR_FH}.  However, independent estimates based on empirical relationships between black hole mass and high-frequency QPOs have suggested values of $M$ between $\sim8.7$ and $15.6$ \msun\ \citep{Rebusco_IGRMass,Iyer_Bayes,Iyer_IGRMass}, while multi-wavelength observations of the hard-to-soft state transition have suggested values of $D$ between $\sim$11 and $\sim$17 kpc \citep{Rodriguez_D}.  This implies that IGR J17091 may have an accretion rate $\dot{m}$ that is significantly below its Eddington rate.
\par The suggestion that IGR J17091 accretes at significantly below the Eddington Limit raises several questions.  Despite evidence of disc winds in IGR J17091 faster than 0.03c \citep{King_IGRWinds}, the wind generation procedure described in \citet{Neilsen_GRSModel} cannot take place without near-Eddington-limited accretion.   Additionally, it makes it increasingly unclear what differentiates IGR J17091 and GRS 1915 from the vast majority of LMXBs which do not display such complex behaviour, and what physical system properties are required for GRS 1915-like variability to be observed.  This latter point was further complicated by the observation of GRS 1915-like variability in 2 out of 155 \rxte\ observations of the Rapid Burster \citep{Bagnoli_RB}.  As thermonuclear (Type I) X-ray bursts are also seen in the Rapid Burster \citep{Hoffman_RB}, it is known to contain a neutron star.  As such, the presence of variability classes in this object rules out any black hole-specific effects as the cause of the complex variability.  In addition to this, the persistent luminosity of the Rapid Burster during periods of GRS 1915-like variability is known to be no greater than $\sim10\%$ of its Eddington limit \citep{Bagnoli_PopStudy}.
\par Accounting for GRS 1915-like variability is required for a complete understanding of the physics of accretion in LMXBs.  As such, \citet{Belloni_GRS_MI} performed a complete model-independent analysis of variability classes in GRS 1915.  This work highlighted the breadth and diversity of variability in GRS 1915, and allowed these authors to search for features common to all variability classes.  For example, \citet{Belloni_GRS_MI} also found that every variability class can be expressed as a pattern of transitions between three quasi-stable phenomenological states.
\par Previous works have noted that some of the variability classes seen in IGR J17091-3624 appear very similar to those seen in GRS 1915 (e.g. \citealp{Altamirano_IGR_FH, Zhang_IGR}).  However, although $\rho$-like classes in the two objects both show lags between hard and soft X-rays photons, these lags appear to possess different signs \citep{Altamirano_IGR_FH}.  Additionally, at least two variability classes have been reported in IGR J17091 which have not yet been reported in GRS 1915 \citep{Pahari_IGRClasses}.  Previous works have described some of the behaviour seen in IGR J17091 in the context of the variability classes described by \citealt{Belloni_GRS_MI} for GRS 1915 (e.g. \citealp{Altamirano_IGR_FH,Pahari_RhoDiff}).  To further explore the comparison between GRS 1915 and IGR J17091, here we perform the first comprehensive model-independent analysis of variability classes in IGR J17091 using the complete set of \rxte\ \citep{Bradt_RXTE} data taken of the 2011-2013 outburst of the object.  We also use data from all other X-ray missions that observed the source during this time to analyse the long-term evolution of the outburst.

\section{Data and Data Analysis}

\label{sec:dex}

\par In this paper, we report data from \rxte , \textit{INTEGRAL}, \textit{Swift}, \textit{Chandra}, \textit{XMM-Newton} and \textit{Suzaku} covering the 2011-2013 outburst of IGR J17091.  Unless stated otherwise, all errors are quoted at the 1$\sigma$ level.
\par
\begin{figure*}
    \includegraphics[width=1.8\columnwidth, trim = {0.75cm 1.0cm 1.0cm 0.8cm},clip]{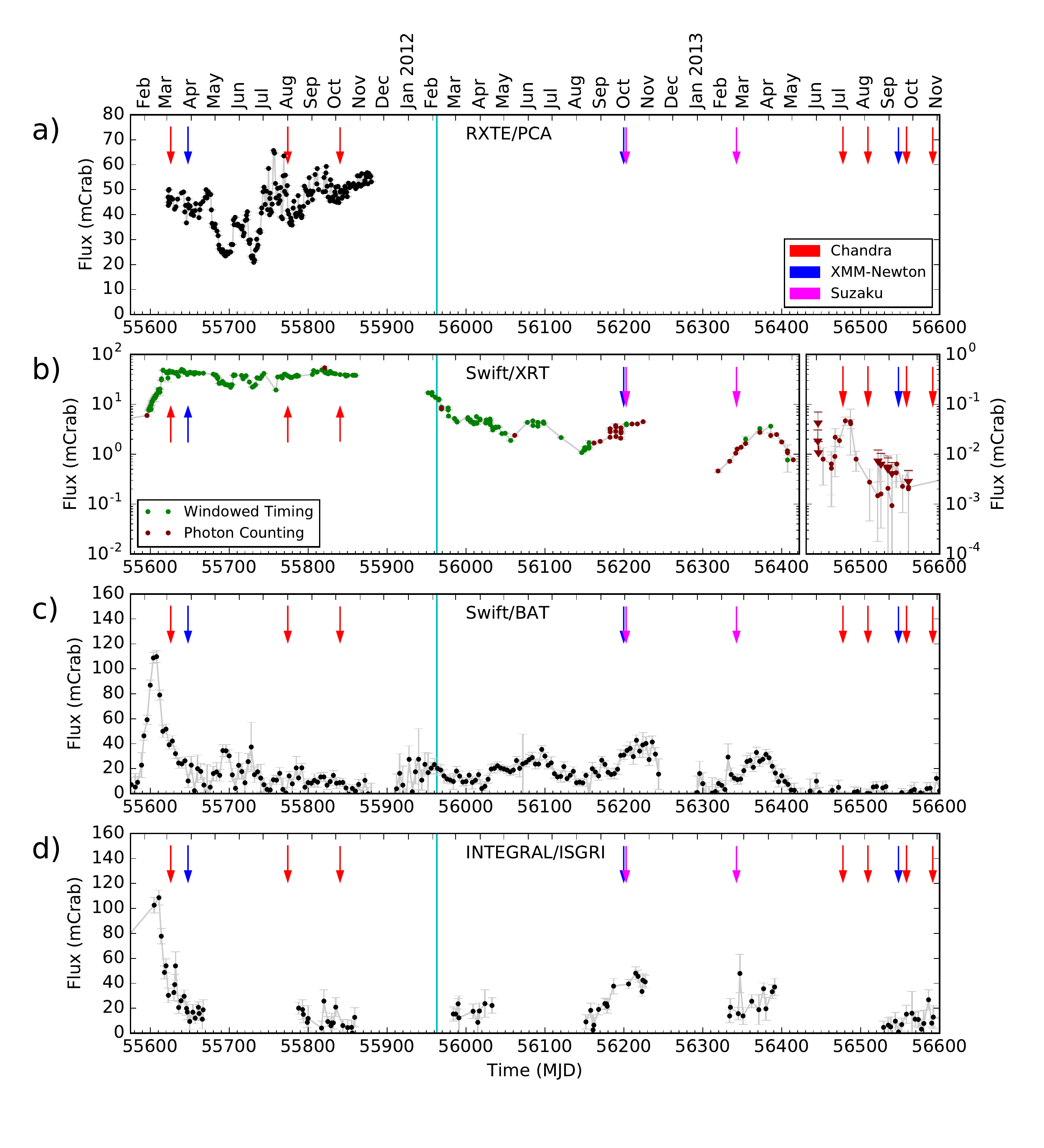}
    \captionsetup{singlelinecheck=off}
    \caption{\rxte\ (Panel a), \textit{Swift/XRT} (Panel b), \textit{Swift/BAT} (Panel b) and \textit{INTEGRAL}/IBIS (Panel d) lightcurves of IGR J17091-3624 during its 2011-2013 outburst.  Arrows mark times at which \textit{XMM-Newton} (blue), \textit{Chandra} (red) or \textit{Suzaku} (magenta) observed IGR J17091-3624.  The cyan line represents MJD 55963, the approximate time IGR J17091-3624 transitions from the soft to the hard state \citep{Drave_Return}.  \rxte \textit{/PCA} \citep{Jahoda_PCA} data are for the 2--16\,keV energy band and taken from \citep{Altamirano_IGR_FH}, \textit{Swift/BAT} \citep{Barthelmy_BAT} data are for 15--50\,keV, \textit{Swift/XRT} \citep{Burrows_XRT}  data are for 0.3--10\,keV and \textit{INTEGRAL/ISGRI} \citep{Ubertini_IBIS} data are for 20-40\,keV.  Note that the data from \textit{Swift/XRT} (Panel B) are shown with a logarithmic $y$-axis to better show the late time progression of the outburst.  Data points are coloured according to the observing mode used.  The \textit{Swift/XRT} data from times later than MJD 56422 are shown to a different scale to better represent the post-outburst evolution of the source.  All data are presented in 1 day bins, except for data from \textit{Swift/BAT} which is presented in 4 day bins.  See also Figure \ref{fig:WhereCls}, in which data from \rxte \textit{/PCA} is presented on a smaller scale.  The Crab count rates used to normalise these data were 2300 cts s$^{-1}$ PCU$^{-1}$, 747.5 cts s$^{-1}$, 0.214 cts s$^{-1}$ and 183.5 cts s$^{-1}$ for \rxte , \textit{Swift/XRT}, \textit{Swift/BAT} and \textit{INTEGRAL/ISGRI} respectively.  \rxte\ data have not been corrected for the 25' offset to avoid contamination from GX 349+2, and for all instruments we implicitly assume that IGR J17091 presents a Crab-like spectrum.}
   \label{fig:allmissions}
\end{figure*}

 In Figure \ref{fig:allmissions} we present long-term lightcurves from \rxte ,  \textit{INTEGRAL} and \textit{Swift} with one datapoint per observation, as well as indicating when during the outburst \textit{Chandra}, \textit{XMM-Newton} and \textit{Suzaku} observations were made.

\subsection{\rxte}

\label{sec:XTEDA}

\par For our variability study, we focus on the data from the Proportional Counter Array (\textit{PCA}, \citealp{Jahoda_PCA}) aboard the Rossi X-Ray Riming Experiment (\rxte , \citealp{Bradt_RXTE}).  We analysed all 2011 \textit{PCA} observations of IGR J17091, corresponding to OBSIDs 96065-03, 96103-01 and 96420-01.  The observations taken for proposals 96065-03 and 96103-01 were contaminated by the nearby X-ray source GX 349+2 \citep{Altamirano_IGR_FH,Rodriguez_Contamination}.  As such we only use observations performed for proposal 96420-01, corresponding to a total of 243 orbits from 215 separate observations.  These were offset by 25' such that GX 349+2 was not in the $1^\circ$ \textit{PCA} field of view.  \rxte\ was decommissioned during a period of Sun constraint centred on MJD 55907,  and hence the last observation of IGR J17091 was taken on MJD 55879.
\par We extracted data from the native \texttt{FITS} format using our own software\footnote{\url{https://github.com/jmcourt/pantheon}}.  To perform medium- to high-frequency ($\gtrsim1$\,Hz) timing analysis, we merged files formatted in \pca\ 's `Good Xenon' data mode and extracted their data at the maximum time resolution ($\sim9.5\times10^{-7}$ s) without accounting for the background.  We divided these data into 128\,s segments as this allowed us to reach frequencies below $\sim0.015$\,Hz, partly sampling the high amplitude quasi-periodic flaring behaviour seen in many classes.  Using the Fast Fourier Transform (FFT), we produced the power spectrum of each segment separately.  We then averaged these spectra to create a one co-added Power Density Spectrum (PDS) for each observation.
\par For low-frequency ($\leq1$\,Hz) timing and correlated spectral/timing analysis, we rebinned the data to 0.5\,s and normalised count rates by the number of proportional counters (PCUs) active in each observation.  Our choice of 1\,Hz allows us to analyse high amplitude `flaring' behaviour (seen at frequencies $\lesssim0.5$\,Hz) separately from the lower-amplitude behaviour seen at $\gtrsim5$\,Hz.
\par We split the data into three energy bands: A (\textit{PCA} channels 0--14, $\sim2$--$6$\,keV), B (\textit{PCA} channels 15--35, $\sim6$--$16$\,keV) and C (\textit{PCA} channels 36--255, $\sim16$--$60$\,keV).  We chose these energy bands to be consistent with the energy bands used by the model-independent classification of variability classes of GRS 1915 in \citet{Belloni_GRS_MI}.  For each of the energy-filtered lightcurves produced we estimated background using \texttt{pcabackest} from the \texttt{FTOOLS} package \citep{Blackburn_FTools} with the \textit{PCA} faint source background model\footnote{\url{http://heasarc.gsfc.nasa.gov/FTP/xte/calib\_data/pca\_bkgd/Faint/pca\_bkgd\_cmfaintl7\_eMv20051128.mdl}}. In all observations, we found that counts in the C band were consistent with background.  We then created Lightcurves $L_A$ and $L_B$ from background-subtracted photons counted in the A and B bands respectively.  We used these lightcurves to define the full-band lightcurve ($L_T=L_A+L_B$) and the soft colour ($C_1=L_B/L_A$) of each observation.  To complement the Fourier spectra, we also constructed Generalised Lomb-Scargle Periodograms of $L_T$ from each dataset, a modified version of the standard Lomb-Scargle periodogram \citep{Lomb_LombScargle, Scargle_LombScargle} that takes into account errors in the dataset \citep{Irwin_LombScargle}.  Using the Lomb-Scargle periodogram instead of the Fourier periodogram here allows us to sample the low-frequency behaviour of lightcurves with data gaps.  This is important, for example, in lightcurves which show two populations of flares, as it allows each population to be studied independently by cropping the other from the lightcurve.
\par We also used data from \citealt{Altamirano_IGR_FH} to sample the long-term colour evolution of IGR J17091.  We use 2 hardness ratios defined by \citeauthor{Altamirano_IGR_FH}: $H_{A1}$ and $H_{A2}$, corresponding to the ratios of the 2--3.5\,keV band against the 3.5--6\,keV band and the 6--9.7\,keV band against the 9.7--16\,keV band respectively.
\par When possible, if low-frequency peaks were present in the Lomb-Scargle spectrum of an observation, we used the position of the highest peak to define a value for a period.  This period was then used to rebin data by phase (or `fold' the data) to search for reccurent hysteretic patterns in the hardness-Intensity diagram (hereafter HID$_1$, a plot of $L_T$ against $C_1$).  We found that quasi-periodic oscillations in our observations tended to show significant frequency shifts on timescales shorter than the length of the observations.  As such, we employed the variable-period folding algorithm outlined in Appendix \ref{app:Flares} where appropriate.  For cases in which this algorithm was not appropriate, we considered small sections of each lightcurve, with a length equivalent to small number of periods, before performing folding.
\par Additionally, in observations which showed a pattern of high-amplitude X-ray flaring in $L_T$, we used our own algorithm to find individual flares (this algorithm is described in Appendix \ref{app:Flares}) and collect statistics on the amplitude, duration and profile of these events.
\par A list of all observations used in this study can be found in Appendix \ref{app:Obsids}.

\subsection{\textit{Swift}}

\par In this paper, we consider data from the Burst Alert Telescope (\textit{BAT}, \citealt{Barthelmy_BAT}), and the X-ray Telescope (\textit{XRT}, \citealt{Burrows_XRT}) aboard Swift Gamma-ray Burst Mission \citep{Gehrels_Swift}.  IGR J17091-3624 was observed with \textit{XRT} for a total of 172 pointed XRT observations between MJDs 55575 and 56600, corresponding to Target IDs 31921, 34543, 30967, 30973, 31920, 35096, 67137, 81917, 522245, 677582 and 677981.  These observations were interrupted during sun constraints centred on MJDs 55907 and 56272.  We created a long-term 0.3--10\,keV \textit{Swift/XRT} light curve, with one bin per pointed observation, using the online light-curve generator provided by the UK Swift Science Data Centre (UKSSDC; \citealp{Evans_Swift1}).  We have also created a long-term 15--50\,keV lightcurve using the publicly available \textit{Swift/BAT} daily-averaged light curve\footnote{\url{http://swift.gsfc.nasa.gov/results/transients/weak/IGRJ17091-3624/}}.  These are shown in Figure \ref{fig:allmissions} Panels (b) and (c) respectively.

\subsection{\textit{INTEGRAL}}

\par The INTErnational Gamma Ray Astrophisical Laboratory (\textit{INTEGRAL}, \citealp{Winkler_IBIS}) is a medium-sized ESA mission launched in 2002. Unique hard X-ray (15--1000\,keV with the ISGRI detector plane) sensitivity and wide field of view make \textit{INTEGRAL} ideally suited to surveying the hard X-ray sky (see \citet{Bird_Survey} and \citet{Krivonos_Survey} for recent surveys). 
\par \textit{INTEGRAL} observations are divided into short ($\sim$2\,ks) pointings called Science Windows (ScWs). We analyse all available observations of IGR J17091 with \textit{INTEGRAL}/IBIS \citep{Ubertini_IBIS} between MJD 55575--55625 where the source is less than 12 degrees from the centre of the field of view and where there is more the 1\,ks of good ISGRI time per ScW. This corresponds to the spectrally hardest period of the 2011-2013 outburst. The filtering of observations results in a total of 188 Science Windows which were processed using the Offline Science Analysis (OSA) software version 10.2 following standard data reduction procedures\footnote{http://www.isdc.unige.ch/integral/analysis} in four energy bands (20--40, 40--100, 100--150, 150--300\,keV). These bands were selected as they are standard energy bands used in the surveys of \citet{Bird_Survey} and \citet{Bazzano_Survey} and allow comparison to these previous works. Images were created at the ScW level as well as a single mosaic of all Science Windows in each energy band.

\subsection{\textit{XMM-Newton}}
\label{sec:xmmdata}

\par In this paper we only consider data from the European Photon Imaging Camera (\textit{EPIC} \citealp{Bignami_EPIC}) aboard \textit{XMM-Newton} \citep{Jansen_XMM}.   \textit{EPIC} consists of one telescope with a pn-type CCD (\textit{EPIC-pn}, \citealp{Struder_PN}) and two telescopes with MOS CCDs (\textit{EPIC-MOS1} and \textit{-MOS2}, \citealp{Turner_MOS}).
\par \textit{XMM/Newton} observed IGR J17091 thrice during the period from 2011--2013 (represented by the blue arrows in Figure \ref{fig:allmissions}).  One of these (OBSID 0721200101) was made on 12 September 2013; we do not consider this observation further as IGR J17091 had returned to quiescence by this time \citep{Altamirano_Quiescence}.  The remaining two observations, corresponding to OBSIDs 0677980201 and 0700381301 respectively, were taken on March 27 2011 (MJD 55647) and September 29 2012 (MJD 56199).
\par During observation 0677980201, \textit{EPIC-pn} was operating in burst mode and \textit{EPIC-MOS} was operating in timing mode.  Given the low efficiency of burst mode, we only consider data from \textit{EPIC-MOS} for this observation.  During observation 0700381301, \textit{EPIC-pn} was operating in timing mode, and thus we use data from \textit{EPIC-pn} for this observation.
\par We used the \textit{XMM-Newton} Science Analysis Software version 15.0.0 (\texttt{SAS}, see \citealp{Ibarra_sas}) to extract calibrated event lists from \textit{EPIC} in both observations.  We used these to construct lightcurves to study the X-ray variability, following standard analysis threads\footnote{\url{http://www.cosmos.esa.int/web/xmm-newton/sas-threads}}.

\subsection{\textit{Chandra}}

\par In this paper we consider data from the Advanced CCD Imaging Camera (\textit{ACIS}, \citealp{Nousek_ACIS}) and the High Resolution Camera (\textit{HRC}, \citealp{Murray_HRC}) aboard \textit{Chandra} \citep{Weisskopf_Chandra}.  \textit{Chandra} made 7 observations of IGR J17091 during the period 2011--2013.  Four of these observations were taken after IGR J17091 returned to quiescence, and we do not consider these further in this paper.  The Chandra observations log is reported in Table \ref{tab:Chandra}. 

\begin{table}
\centering
\begin{tabular}{lllllll}
\hline
\hline
\scriptsize OBSID &\scriptsize  Instrument &\scriptsize Grating &\scriptsize Exposure (ks) &\scriptsize  Mode &\scriptsize MJD\\
\hline
12505  	& \textit{HRC-I}    &   NONE      &    1.13      & $I$ & 55626\\
12405  	& \textit{ACIS-S} &   HETG     &    31.21     & $C$ & 55774\\
12406  	& \textit{ACIS-S} &   HETG     &    27.29     & $T$ & 55840\\
\hline
\hline
\end{tabular}
\caption{Chandra observations log covering the three observations considered in this paper.  $I$ refers to Imaging mode, $C$ refers to CC33\_Graded mode and $T$ refers to Timed Exposure Faint mode.  HETG refers to the High Energy Transmission Grating.}
\label{tab:Chandra}
\end{table}

\par We analysed these data using \texttt{CIAO} version 4.8 \citep{Fruscione_Ciao}, following the standard analysis threads. In order to apply the most recent calibration files (CALDB 4.7.0, \citealp{Graessle_ChaCALDB}), we reprocessed the data from the three observations using the \texttt{chandra\_repro} script\footnote{See e.g. \url{http://cxc.harvard.edu/ciao/ahelp/chandra_repro.html}}, and used this to produce data products following standard procedures.
\par The first Chandra observation(OBSID 12505) of this source was made shortly after it went into outburst in February 2011. It was a 1\,ks observation performed to refine the position of the X-Ray source, using the High-Resolution Camera in Imaging mode (HRC-I). We created the 0.06--10\,kev light curve accounting for the Dead Time Factor (DTF) to correct the exposure time and count rate, due to the deviation of the detector from the standard detection efficiency,  using the \texttt{dmextract} tool in the \texttt{CIAO} software.
\par Two additional observations (OBSIDs 12405 and 12406) were performed within 214 days of this first observation, using the High Energy Transmition Grating Spectrometer (HETGS) on board \textit{Chandra}. The incident X-Ray flux was dispersed onto \textit{ACIS} using a narrow strip array configuration (ACIS-S). Continuous Clocking and Time Exposure modes were use in each observation respectively (see \citealp{King_IGRWinds} for further details). We exclude any events below 0.4\,keV, since the grating efficiency is essentially zero below this energy. In the case of the OBSID 12405 observations we also excluded the Flight Grade 66 events in the event file, as they were not appropriately graded. We extracted the 0.5-10\,kev HEGTS light curves, excluding the zeroth-order flux, adopting standard procedures.

\subsection{\textit{Suzaku}}

\par In this paper, we only consider data from the X-ray Imaging Spectrometer (\textit{XIS}, \citealp{Koyama_XIS}) aboard \textit{Suzaku} \citep{Mitsuda_Suzaku}.  \textit{Suzaku} observed IGR J17091 twice during the period 2011--2013; a 42.1\,ks observation on October 2--3, 2012 (MJD 56202--56203, ObsID: 407037010) and an 81.9\,ks observation on February 19--21, 2013 (MJD 56342--56344, ObsID: 407037020). \textit{XIS} consists of four X-ray CCDs (\textit{XIS} 0, 1, 2 and 3), and all them except for XIS 2 were operating in the 1/4 window mode which has a minimum time resolution of 2 seconds.
\par We analysed the \textit{Suzaku} data using {\it HEASOFT} 6.19 in the following standard procedures after reprocessing the data with \texttt{aepipeline} and the latest calibration database (version 20160607).  We extracted \textit{XIS} light curves in the 0.7--10 keV range, and subtracted background individually for XIS 0, 1 and 3 and then summed these to obtain the total background.  We created Power density spectra (PDS) using {\tt powspec} in the {\tt XRONOS} package.

\section{Results}
\label{sec:results}

\subsection{Outburst Evolution}

\par The 2011-2013 outburst of IGR J17091-3624 was first detected with \textit{Swift/XRT} on MJD 55595 (3 Feb 2011) \citep{Krimm_IGROutburst}, and was observed by \rxte\ , \textit{Swift/XRT}, \textit{Swift/BAT} and \textit{INTEGRAL/ISGRI} (see Figure \ref{fig:allmissions}, Panels a, b, c, d respectively).  There were also pointed observations by \textit{XMM-Newton}, \textit{Chandra} and \textit{Suzaku} during this time (denoted by coloured arrows in Figure \ref{fig:allmissions}).
\par \rxte\ data were taken within the first week of the outburst, but they were heavily contiminated by the nearby source GX 349+2 \citep{Rodriguez_Contamination}.  Thus these data are not considered here.
\par The onset of the outburst can be seen in the \textit{Swift/BAT} lightcurve (Figure \ref{fig:allmissions} Panel c).  In a 22 day period between MJDs 55584 and 55608, the 15--50\,keV intensity from IGR J17091 rose from $\sim9$\,mCrab to a peak of $\sim110$\,mCrab.  This onset rise in intensity can also be seen in 0.3--10\,keV \textit{Swift/XRT} data and 20--40\,keV \textit{INTEGRAL/ISGRI} data.
\par After peak intensity, the 15--50\,keV flux (\textit{Swift/BAT}) began to steadily decrease, until returning to a level of $\sim$20\,mCrab by MJD 55633.  A similar decrease in flux can be seen in the data obtained by \textit{INTEGRAL} at this time (Figure \ref{fig:allmissions} Panel (d).  However, there was no corresponding fall in the flux at lower energies; both the long-term 2--16\,keV \rxte\ data and \textit{Swift/XRT} data (Panels a and b respectively) show relatively constant fluxes of 45\,mCrab between MJDs 55608 and 55633.
\par The significant decrease in high-energy flux during this time corresponds to IGR J17091 transitioning from a hard state to a soft intermediate state \citep{Pahari_RhoDiff}.  This transition coincides with a radio flare reported by \citet{Rodriguez_D} which was observed by the Australian Telescope Compact Array (\textit{ATCA}).
\par \citealp{Altamirano_10Hz} first reported a 10\,mHz QPO in \rxte\ data on MJD 55634 , evolving into `Heartbeat-like' flaring by MJD 55639 \citep{Altamirano_Discovery}.  Between MJDs 55634 and 55879, the global \rxte\ lightcurve shows large fluctuations in intensity on timescales of days to weeks, ranging from a minimum of $\sim1$\,mCrab on MJD 55731 to a maximum of $\sim66$\,mCrab on MJD 55756.  The \textit{Swift/XRT} lightcurve shows fluctuations that mirror those seen by \rxte\ during this period, but the amplitude of the fluctuations is significantly reduced.
\par \textit{Swift/XRT} was unable to observe again until MJD 55952.  Between this date and MJD 55989, \textit{Swift/XRT} observed a gradual decrease in intensity corresponding to a return to the low/hard state \citep{Drave_Return}.
\par Between MJD 55989 and the end of the outburst on MJD 56445, we see secondary peaks in the \textit{Swift/XRT}, \textit{Swift/BAT} and \textit{INTEGRAL/ISGRI} lightcurves that evolve over timescales of $\lesssim100$ days.  Similar humps have been seen before in lightcurves from other objects, for example the black hole candidate XTE J1650-500 \citep{Tomsick_MiniOutbursts} and the neutron stars SAX J1808.4-3658 \citep{Wijnands_1808} and SAX J1750.8-2900 \citep{Allen_1750}.  These humps are referred to as `re-flares' (also as `rebrightenings',  `echo-outbursts', `mini-outbursts' or a `flaring tail', e.g. \citealp{Patruno_Reflares2}).  We see a total of 3 apparent re-flares in the \textit{Swift/BAT} data, centred approximately at MJDs 56100, 56220 and 56375.
\par The observation with \textit{XMM-Newton/EPIC-pn} on MJD 56547 (12 September 2013) recorded a rate of 0.019 cts s$^{-1}$.  An observation with \textit{EPIC-pn} in 2007, while IGR J17091 was in quiescence \citep{Wijnands_Quiescence}, detected a similar count rate of 0.020 cts s$^{-1}$.  Therefore we define MJD 56547 as the upper limit on the endpoint of the 2011-2013 outburst.  As such the outburst, as defined here, lasted for $\lesssim$952 days.
\par After the end of the 2011-2013 outburst, IGR J17091 remained in quiescence until the start of a new outburst around MJD 57444 (26 February 2016, \citealp{Miller_2016Outburst}).

\subsection{\rxte}

\par Using the data products described in Section \ref{sec:dex}, we assigned a model-independent variability class to each of the 243 \rxte\textit{/PCA} orbits.  To avoid bias, this was done without reference to the classes defined by \citet{Belloni_GRS_MI} to describe the behaviour of GRS 1915.
\par Classes were initially assigned based on analysis of lightcurve profiles, count rate, mean fractional RMS \citep{Vaughan_RMS}, Fourier and Lomb-scargle power spectra and hardness-intensity diagrams.  For observations with significant quasi-periodic variability at a frequency lower than $\sim1$\,Hz, we also attempted to fold lightcurves to analyse count rate and colour as a function of phase.  When flares were present in the lightcurve, we used our algorithm (described in Appendix \ref{app:Flares}) to sample the distribution of parameters such as peak flare count rate, flare rise time and flare fall time.  All parameters were normalised per active PCU, and fractional RMS values were taken from 2--60\,keV lightcurves binned to 0.5\,s.  We identify nine distinct classes, labelled I to IX; we describe these in the following sections.
\par Although the criteria for assigning each class to an observation was different, a number of criteria were given the most weight.  In particular, the detection, $q$-value and peak frequency of a QPO in the range 2\,Hz--10\,Hz were used as criteria for all classes, as well as the presence or absence of high-amplitude quasi-periodic flaring with a frequency between 0.01--1\,Hz.  The folded profile of these flares, as well as the presence of associated harmonics, were also used as classification diagnostics in observations.  Additionally, the presence or absence of low count-rate 'dips' in a lightcurve was used as a criterion for Classes VI, VIII and IX.  Detailed criteria for each individual class are given below in Sections \ref{sec:ClassI} to \ref{sec:ClassIX}.
\par For hardness-intensity diagrams, we describe looping behaviour with the terms `clockwise' and `anticlockwise'; in all cases, these terms refer to the direction of a loop plotted in a hardness-intensity diagram with colour on the $x$-axis and intensity on the $y$-axis.
\par In Appendix \ref{app:Obsids}, we present a list of all orbits used in the study along with the variability classes we assigned to them.
\par
\begin{figure*}
    \includegraphics[width=1.8\columnwidth, trim = {1.3cm 2.0cm 1.8cm 1.8cm},clip]{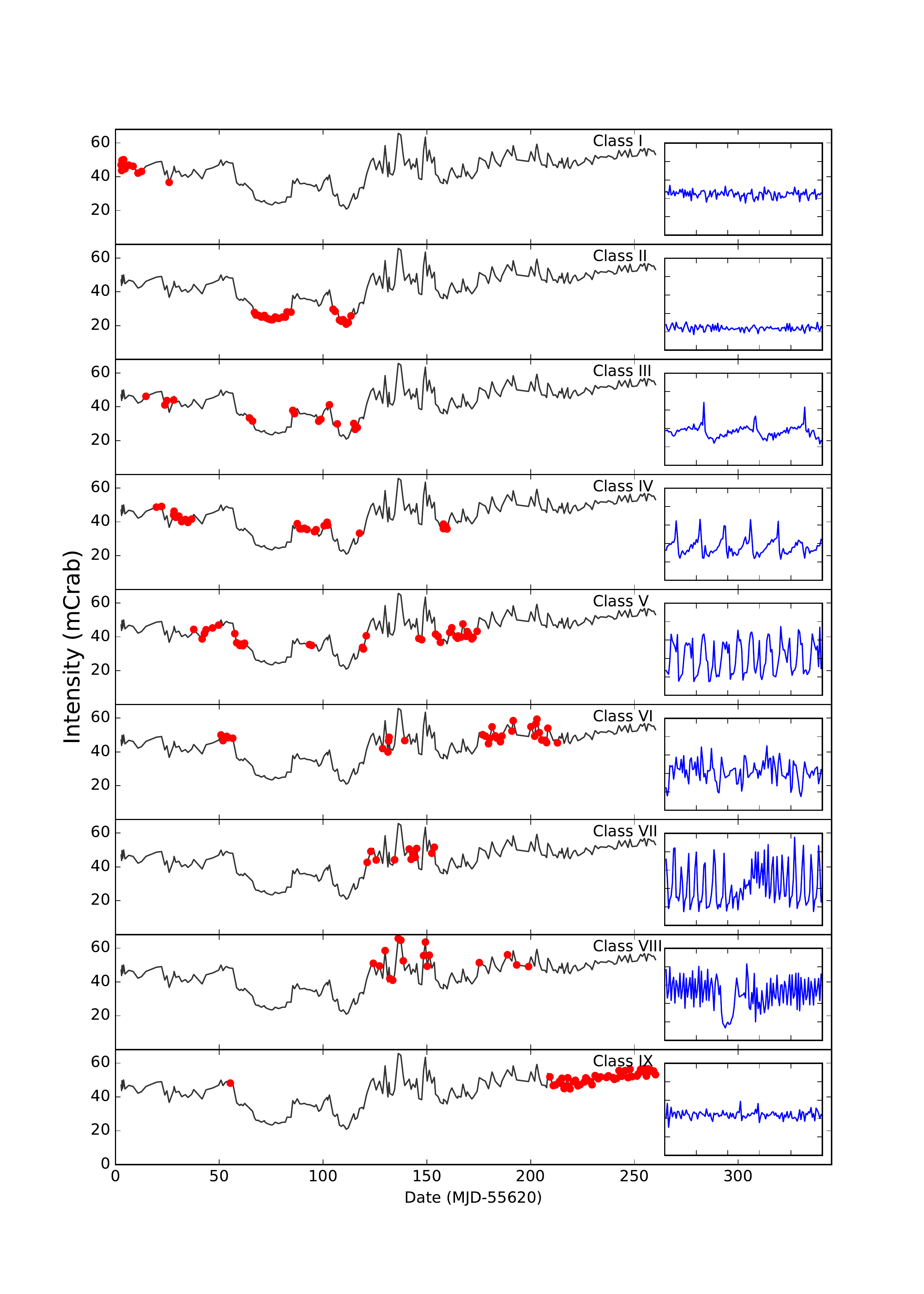}
    \captionsetup{singlelinecheck=off}
    \caption{Global 2--3.5\,keV Lightcurves of IGR J17091-3524 during the 2011-2013 outburst, with each point corresponding to the mean Crab-normalised count rate of a single \rxte\ observation of the object (in turn corresponding to between 0.4 and 3.6 ks of data).  In each lightcurve, every observation identified as belonging to a particular class (indicated on the plot) is highlighted.  These are presented along with a characteristic lightcurve (inset) from an observation belonging to the relevant class.  Each lightcurve is 250\,s in length, and has a $y$-scale from 0 to 250\spcu .  Data taken from \citealt{Altamirano_IGR_FH}.}
   \label{fig:WhereCls}
\end{figure*}

\begin{figure}
\subfloat[\textit{Colour-Colour Diagram}]{\includegraphics[width=\columnwidth, trim = 0mm 0mm 0mm 8mm,clip]{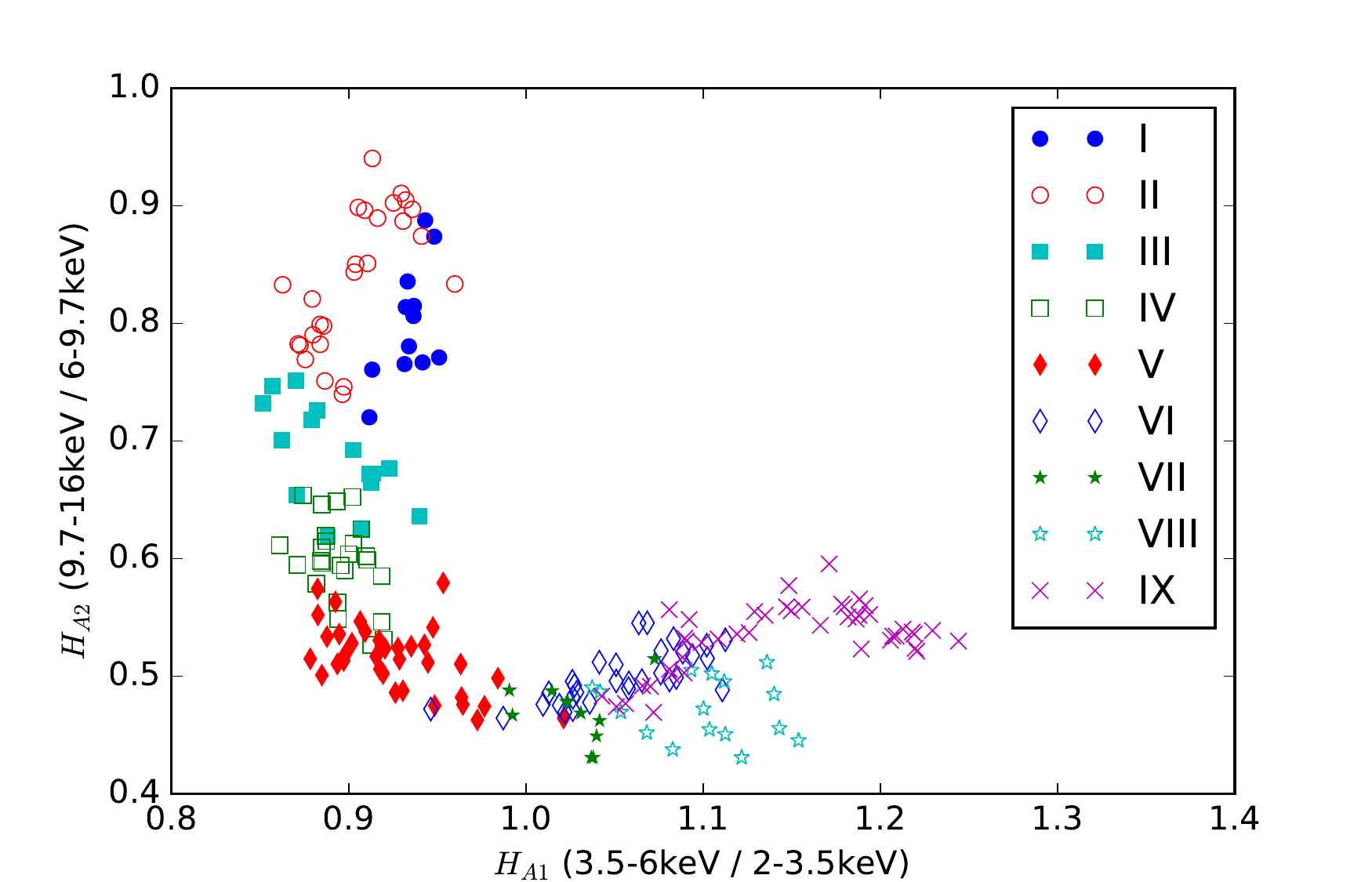}}\\
\subfloat[\textit{"Soft" ($H_{A1}$) Hardness-Intensity Diagram}]{\includegraphics[width=\columnwidth, trim = 0mm 0mm 0mm 8mm,clip]{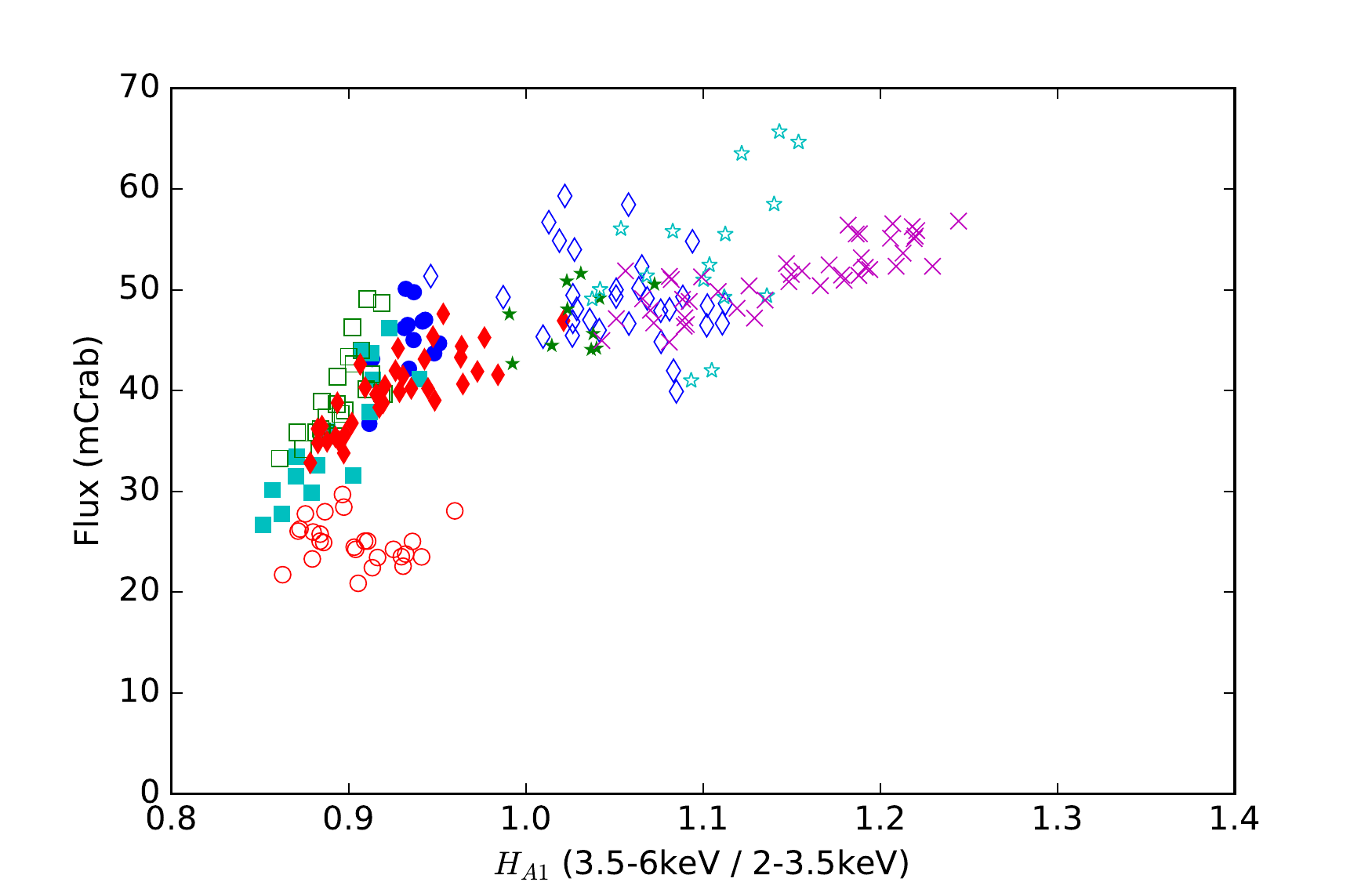}}\\
\subfloat[\textit{"Hard" ($H_{A2}$) Hardness-Intensity Diagram}]{\includegraphics[width=\columnwidth, trim = 0mm 0mm 0mm 8mm,clip]{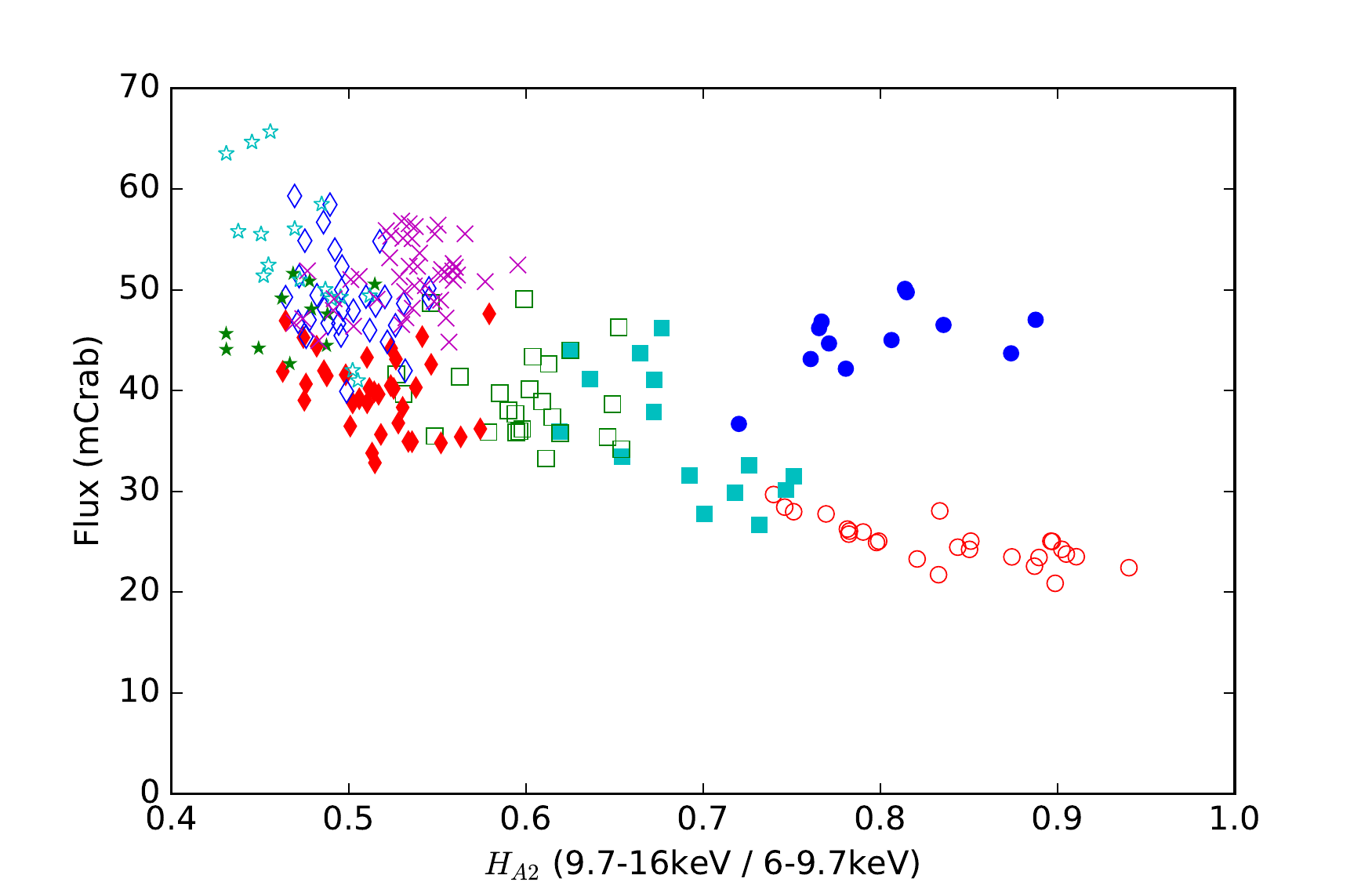}}\\
\captionsetup{singlelinecheck=off}
\caption{A global colour-colour diagram (a), "soft" hardness-intensity diagram (b) and "hard" hardness-intensity diagram (c) of the 2011-2013 outburst of IGR J17091, using the colours $H_{A1}$ and $H_{A2}$ defined previously.  Observations belonging to different classes have been highlighted in different colours.  Data taken from \citealt{Altamirano_IGR_FH}.}
\label{fig:IIIisHarder}
\end{figure}

 In Figure \ref{fig:WhereCls}, we show global 2--16\,keV lightcurves of IGR J17091 during the 2011-2013 outburst.  In each panel, all observations of a given class are highlighted in red.  A characteristic lightcurve is also presented for each class.  In Figure \ref{fig:IIIisHarder} panel (a), we show a plot of average hardness $H_{A2}$ against $H_{A1}$ for each observation, showing the long-term hysteresis of the object in colour-colour space.  Again, observations belonging to each variability class are highlighted.  In Figure \ref{fig:IIIisHarder} panels (b) and (c), we show global hardness-intensity diagrams for $H_{A1}$ and $H_{A2}$ respectively.
\par In Figure \ref{fig:IIIisHarder} Panel (a), we see that IGR J17091-3624 traces a two branched pattern in colour-colour space corresponding to a branch which is soft ($\sim0.9$) in $H_{A1}$ and variable in $H_{A2}$ and a branch which is soft ($\sim0.5$) in $H_{A2}$ and variable in $H_{A1}$.  The `soft' HID shown in Figure \ref{fig:IIIisHarder} Panel (b) is dominated by a branch with a wide spread in $H_{A1}$ and intensities between $\sim40\mbox{--}60$\,mCrab.  A second branch exists at lower intensities, and shows an anticorrelation between intensity and $H_{A1}$.  Finally, the `hard' HID shown in Figure \ref{fig:IIIisHarder} Panel (c) shows an obvious anticorrelation between $H_{A2}$ and intensity, but there is also a secondary branch between $H_{A2}\approx 0.7\mbox{--}0.9$ at a constant intensity of $\sim40$\,mCrab.

\par For characteristic count rates and colours in each class, we quote the upper and lower quartile values \citep{Kenney_Quartile} instead of the mean.  This is due to the presence of high-amplitude but short-lived flares in many of the classes we describe.  Using the upper and lower quartiles as our measure of average and distribution means that our values will be less susceptible to outlier values of count rate and colour present in these flares.  All count rates have been background corrected (see Section \ref{sec:XTEDA}).
\par We have obtained mean values for these count rate quartiles, as well as values for colour $C_1$ and fractional RMS, by calculating these values individually for each orbit.  Histograms were then constructed from these datasets for each class, such that the mean and standard deviation of these values could be measured for each class.  These values are presented in Table \ref{tab:basicparams}.
\par We describe QPOs in terms of their $q$-value; a measure of coherence defined by the ratio of peak frequency and full-width half-maximum of each QPO.  We collected these values by fitting our power spectra with Lorentzians.

\begin{table}
\centering
\begin{tabular}{rllll} 
\hline
\hline
\scriptsize Class &\scriptsize LQ Rate &\scriptsize  UQ Rate &\scriptsize Frac. RMS &\scriptsize Median C$_1$\\
\scriptsize &\scriptsize (cts s$^{-1}$) &\scriptsize (cts s$^{-1}$) & & \\
\hline
I&84--108&106--132&0.13--0.19&0.4--0.68\\
II&43--57&59--71&0.15--0.23&0.4--0.68\\
III&64--84&80--110&0.17-0.23&0.35--0.45\\
IV&63--81&92--122&0.27--0.37&0.32--0.4\\
V&49--67&88--134&0.44--0.54&0.28--0.46\\
VI&64--98&111--155&0.29--0.47&0.33--0.61\\
VII&65--79&128--140&0.45--0.57&0.32--0.42\\
VIII&62--88&142--178&0.42--0.52&0.36--0.49\\
IX&87--111&114--144&0.16--0.24&0.42-0.6\\
\hline
\hline
\end{tabular}
\caption{Lower and upper quartile count rates, fractional RMS and median colour averaged across all observations belonging to each class.  Count rates and fractional RMS are taken from the full energy range of \rxte\textit{/PCA}, and fractional RMS values are 2--60\,keV taken from lightcurves binned to 0.5\,s.  Count rates are normalised for the number of PCUs active during each observation.  All values are quoted as $1\sigma$ ranges.}
\label{tab:basicparams}
\end{table}

\par For each class, we present three standard data products; a 500\,s lightcurve, a variable-length lightcurve where the length has been selected to best display the variability associated with the class and a Fourier PDS.  Unless otherwise stated in the figure caption, the 500\,s lightcurve and the Fourier PDS are presented at the same scale for all classes.  In Table \ref{tab:CPopD} we present a tally of the number of times we assigned each Variability Class to an \rxte\ orbit.

\begin{table}
\centering
\begin{tabular}{llll}
\hline
\hline
\scriptsize Class &\scriptsize  Orbits &\scriptsize Total Time (s) &\scriptsize Fraction \\
\hline
I & 31 &  69569 & 14.8\%\\
II & 26 &  50875 & 10.8\%\\
III & 14 &  26228 & 5.6\%\\
IV & 31 &  69926 & 14.9\%\\
V & 35 &  72044 & 15.3\%\\
VI & 29 &  54171 & 11.5\%\\
VII & 11 &  19241 & 4.1\%\\
VIII & 16 &  26553 & 5.7\%\\
IX & 50 &  81037 & 17.3\%\\
\hline
\hline
\end{tabular}
\caption{A tally of the number of times we assigned each of our nine Variability Classes to an \rxte\ orbit.  We have also calculated the amount of observation time corresponding to each class, and thus inferred the fraction of the time that IGR J17091 spent in each class.  Note: the values in the Total Time column assume that each orbit only corresponds to a single variability Class.}
\label{tab:CPopD}
\end{table}

\begin{figure}
    \includegraphics[width=0.8\columnwidth, trim = 0.6cm 0 3.9cm 0]{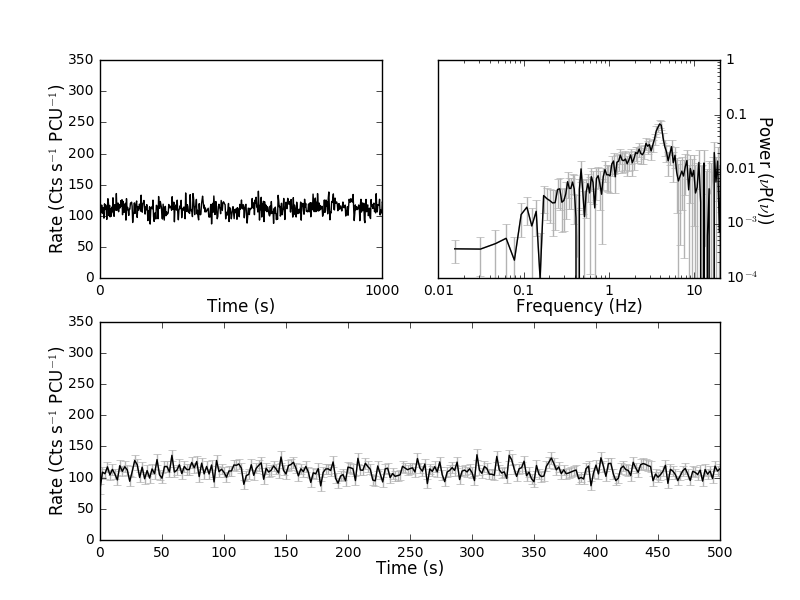}
    \captionsetup{singlelinecheck=off}
    \caption{Plots of the Class I observation 96420-01-01-00, orbit 0.  \textit{Top-left}: 1000\,s lightcurve binned on 2 seconds to show lightcurve evolution.  \textit{Top-right}: Fourier Power Density Spectrum.  \textit{Bottom}: 500\,s lightcurve binned on 2 seconds.}
   \label{fig:Bmulti}
\end{figure}

\subsubsection{Class I --  Figure \ref{fig:Bmulti}}
\label{sec:ClassI}

In the 2\,s binned lightcurve of a Class I observation, there is no structured second-to-minute scale variability.  The Fourier PDS of all observations in this class shows broad band noise between $\sim1$--$10$\,Hz, as well as a weak QPO (with a $q$-value of $\sim5$) which peaks at around 5\,Hz.

\begin{figure}
    \includegraphics[width=0.8\columnwidth, trim = 0.6cm 0 3.9cm 0]{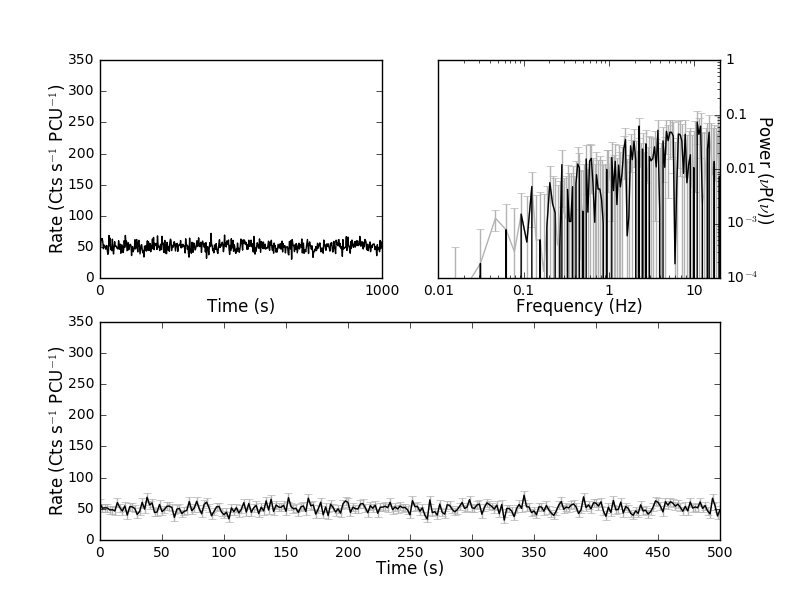}
    \captionsetup{singlelinecheck=off}
    \caption{Plots of the Class II observation 96420-01-11-00, orbit 0.  \textit{Top-left}:  1000\,s lightcurve binned on 2 seconds to show lightcurve evolution.  \textit{Top-right}: Fourier Power Density Spectrum.  \textit{Bottom}: Lightcurve binned on 2 seconds.}
   \label{fig:Emulti}
\end{figure}

\subsubsection{Class II -- Figure \ref{fig:Emulti}}

\par Class II observations are a factor of $\sim2$ fainter in the $L_T$ band than Class I observations.  They also occupy a different branch in a plot of hardness $H_{A2}$ against flux (see Figure \ref{fig:IIIisHarder}c).  The PDS shows no significant broad band noise above $\sim1$Hz unlike that which is seen in Class I.  The $\sim$5Hz QPO seen in Class I is absent in Class II.

\begin{figure}
    \includegraphics[width=0.8\columnwidth, trim = 0.6cm 0 3.9cm 0]{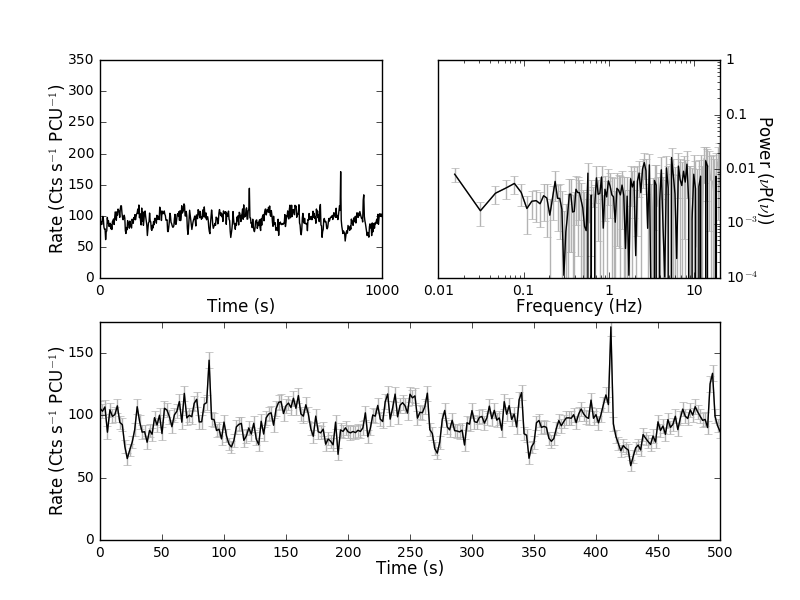}
    \captionsetup{singlelinecheck=off}
    \caption{Plots of the Class III observation 96420-01-04-01, orbit 0.  \textit{Top-left}: 1000\,s lightcurve binned on 2 seconds to show lightcurve evolution.  \textit{Top-right}: Fourier Power Density Spectrum.  \textit{Bottom}: Lightcurve binned on 2 seconds.  Note that, to emphasise the behaviour of the lightcurve in this class, we have magnified the 500\,s lightcurve y-scale by a factor of 2 compared with the lightcurves presented for other classes.}
   \label{fig:Gmulti}
\end{figure}

\subsubsection{Class III -- Figure \ref{fig:Gmulti}}
\label{sec:classIII}

\par Unlike Classes I \& II, Class III lightcurves show structured flaring, with a peak-to-peak recurrence time of $42$--$80$\,s.  Most flares consist of a steady $\sim60$\,s rise in count rate and then an additional and sudden rise to a peak count rate at $\gtrsim200$\spcu which lasts for $\lesssim$0.5\,s before returning to continuum level (we have magnified the y-scaling in the lightcurve of Figure \ref{fig:Gmulti} to emphasise this behaviour). This sudden rise is not present in every flare; in some observations it is absent from every flare feature.  No 5Hz QPO is present in the PDS and there is no significant variability in the range between $\sim1\mbox{--}10$Hz.

\par
\begin{figure}
    \includegraphics[width=\columnwidth, trim = 0mm 0mm 0mm 0mm]{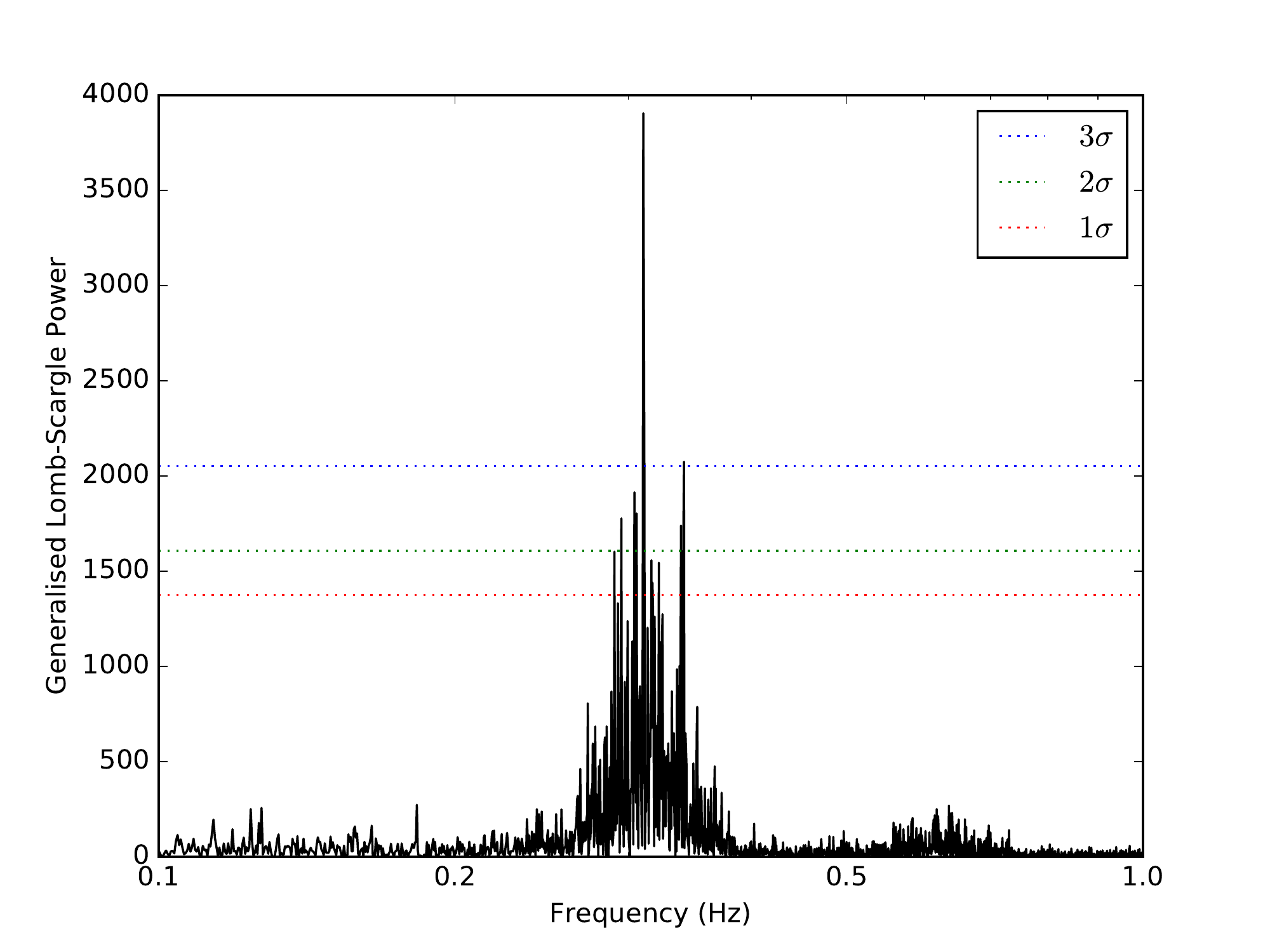}
    \captionsetup{singlelinecheck=off}
    \caption{The Lomb-Scargle periodogram of observation 96420-19-01, orbit 0, with significance levels of 1, 2 and 3$\sigma$ plotted.  The peak at 0.31\,Hz was used to define a QPO frequency when folding the data from this observation.}
   \label{fig:IIILS}
\end{figure}

\begin{figure}
    \includegraphics[width=\columnwidth, trim = 0mm 0mm 0mm 0mm]{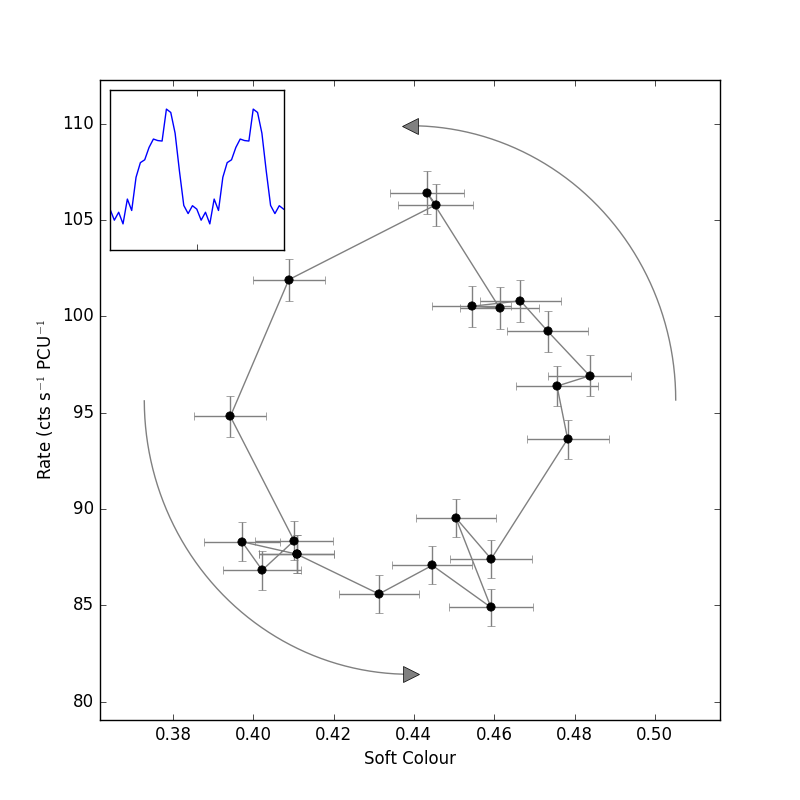}
    \captionsetup{singlelinecheck=off}
    \caption{The hardness-intensity diagram (HID$_1$) of the Class III observation 96420-01-04-01, orbit 0.  The data have been folded over a period of 79.61 s, corresponding to the peak frequency in the Lomb-Scargle spectrum of this observation.  Inset is the folded lightcurve of the same data.}
   \label{fig:LoopIII}
\end{figure}

 As this class has a well-defined periodicity, we folded data in each observation to improve statistics using the best-fit period obtained from generalised Lomb-Scargle Periodogram Analysis; we show a representative Lomb-Scargle periodogram in Figure \ref{fig:IIILS}.  We find an anticlockwise hysteretic loop in the folded HID$_1$ of all 15 Class III orbits.  In Figure \ref{fig:LoopIII} we show an example of one of these loops.

\begin{figure}
    \includegraphics[width=0.8\columnwidth, trim = 0.6cm 0 3.9cm 0]{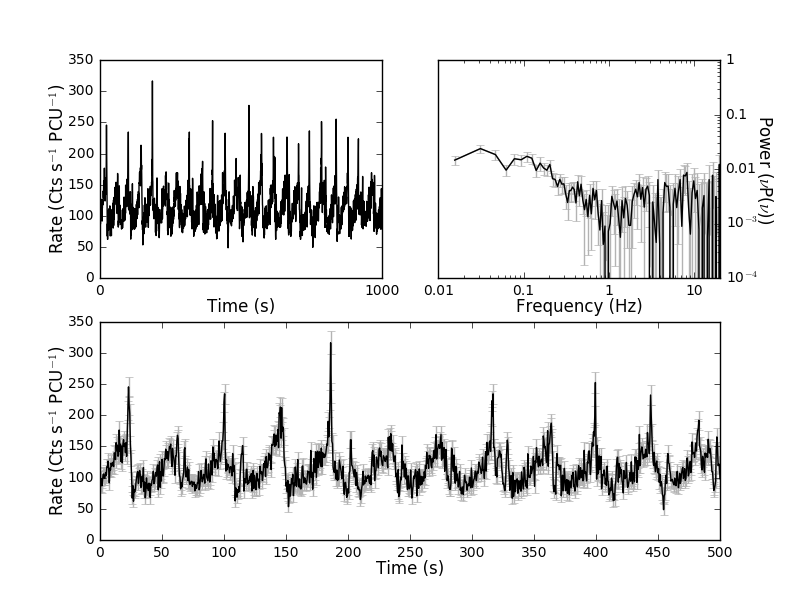}\\
    \captionsetup{singlelinecheck=off}
    \caption{Plots of the Class IV observation 96420-01-05-00, orbit 0.  \textit{Top-left}: 1000\,s lightcurve binned on 2 seconds to show lightcurve evolution.  \textit{Top-right}: Fourier Power Density Spectrum.  \textit{Bottom}: Lightcurve binned on 0.5 seconds.}
   \label{fig:Jmulti}
\end{figure}

\subsubsection{Class IV -- Figure \ref{fig:Jmulti}}
\label{sec:classIV}

\par The lightcurves in this class show regular variability with a peak-to-peak recurrence time of $25$--$39$\,s.  We performed peak analysis (see Appendix \ref{app:Flares}) on observations belonging to this class, finding that each peak has a rise time with lower and upper quartile values of $19.5$ and $33.5$ s, a fall time with lower and upper quartile values of $4.6$ and $13.5$\,s and a peak count rate of $159$--$241$\spcu\ .  There are no prominent significant QPOs in the Fourier PDS above $\sim1$Hz.
\par
\begin{figure}
    \includegraphics[width=\columnwidth, trim = 0mm 0mm 0mm 0mm]{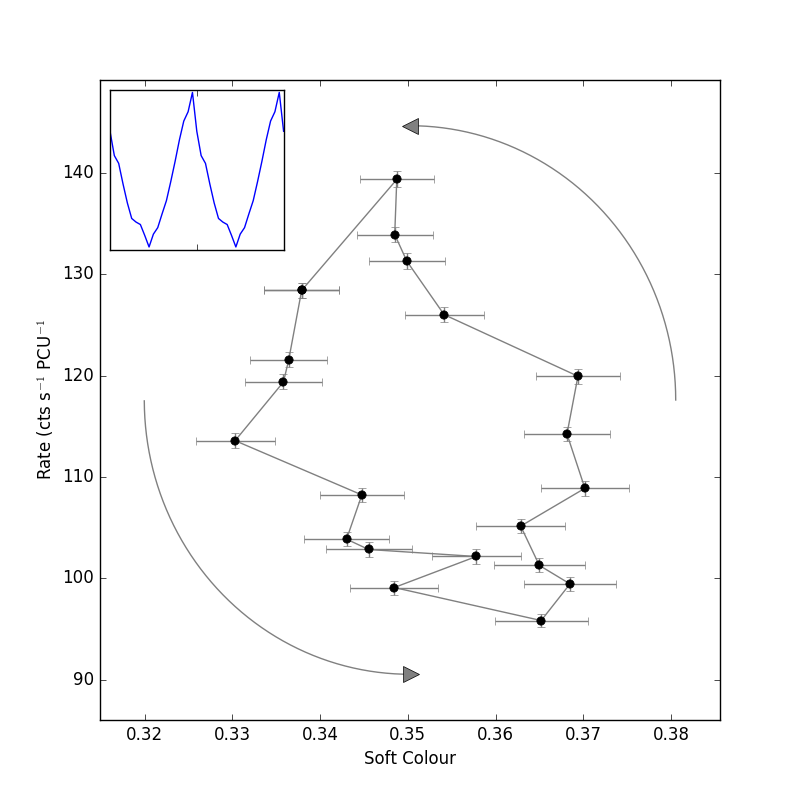}\\
    \includegraphics[width=0.5\columnwidth, trim = 0mm 0mm 0mm 0mm]{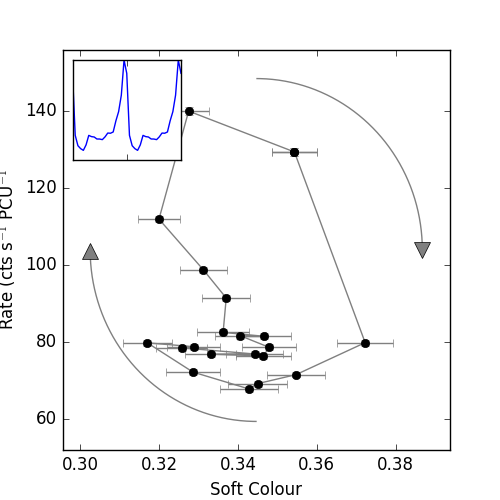}\includegraphics[width=0.5\columnwidth, trim = 0mm 0mm 0mm 0mm]{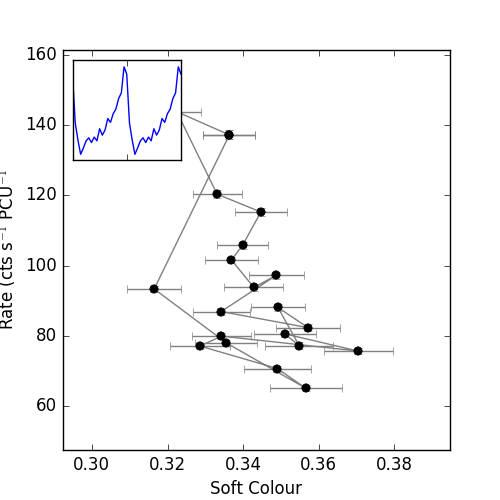}
    \captionsetup{singlelinecheck=off}
    \caption{\textit{Top}: The hardness-intensity diagram (HID$_1$) of the Class IV observation 96420-01-05-00, orbit 0 showing an anticlockwise loop.  The data have been folded over a variable period found with the algorithm described in Appendix \ref{app:Flares}.  Inset is the folded lightcurve of the same data.  \textit{Bottom Left}: The hardness-intensity diagram of Class IV observations 96420-01-24-02 orbit 0, an example of a clockwise loop.  \textit{Bottom Right}: The hardness-intensity diagram of Class IV observation 96420-01-06-00 orbit 0, in which we were unable to ascertain the presence of a loop.}
   \label{fig:LoopIV}
\end{figure}

 We folded individual Class IV lightcurves and found anticlockwise hysteretic loops in the HID$_1$ of 14 out of 30 Class IV observations.  In the top panel of Figure \ref{fig:LoopIV} we show an example of one of these loops.  However, we also find clockwise hysteretic loops in 6 Class IV observations, and in 10 orbits the data did not allow us to ascertain the presence of a loop.  We provide an example of both of these in the lower panels of Figure \ref{fig:LoopIV}.  We note that the structure of clockwise loops are more complex than anticlockwise loops in Class IV, consisting of several lobes\footnote{In HIDs with multiple lobes, the loop direction we assign to the observation corresponds to the direction of the largest lobe.} rather than a single loop (Figure \ref{fig:LoopIV}, bottom-left).

\par Compared with Class III, the oscillations in Class IV occur with a significantly lower period, with a mean peak-to-peak recurrence time of $\sim30$\,s compared to $\sim60$\,s in Class III.
\par In Figure \ref{fig:IIIisHarder} we show that Classes III and IV can also be distinguished by average hardness, as Class III tends to have a greater value of $H_{A2}$ than Class IV.

\begin{figure}
    \includegraphics[width=0.8\columnwidth, trim = 0.6cm 0 3.9cm 0]{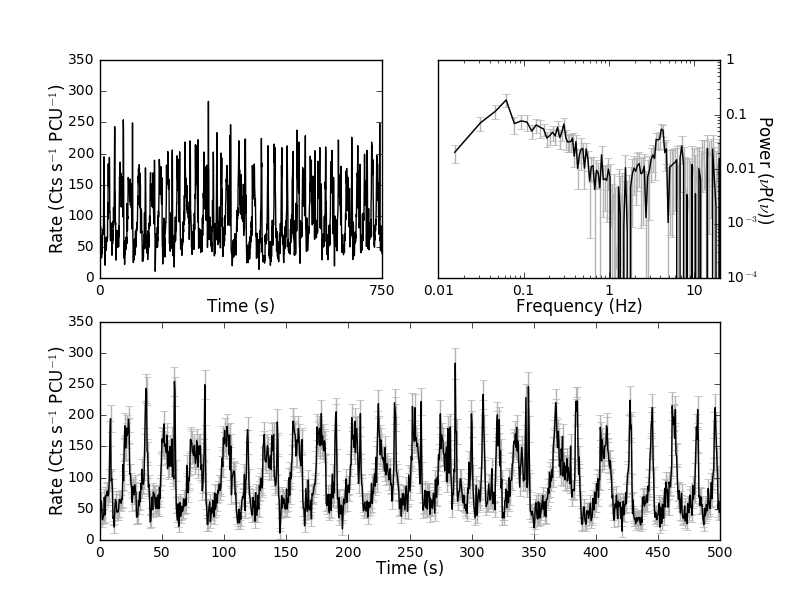}
    \captionsetup{singlelinecheck=off}
    \caption{Plots of the Class V observation 96420-01-06-03, orbit 0.  \textit{Top-left}: 750\,s lightcurve binned on 2 seconds to show lightcurve evolution.  \textit{Top-right}: Fourier Power Density Spectrum. \textit{Bottom}: Lightcurve binned on 0.5 seconds.}
   \label{fig:Kmulti}
\end{figure}

\subsubsection{Class V -- Figure \ref{fig:Kmulti}}
\label{sec:classV}

\par The lightcurves in this class, like in Classes III and IV, show flaring behaviour, with flares separated by a few tens of seconds.  At higher frequencies, the PDS shows a prominent QPO centred at $\sim4$Hz with as $q$-value of $\sim3$.  There is also significant broad band noise between $\sim0.1$--$1$Hz
\par
\begin{figure}
    \includegraphics[width=\columnwidth, trim = 0mm 0mm 0mm 0mm]{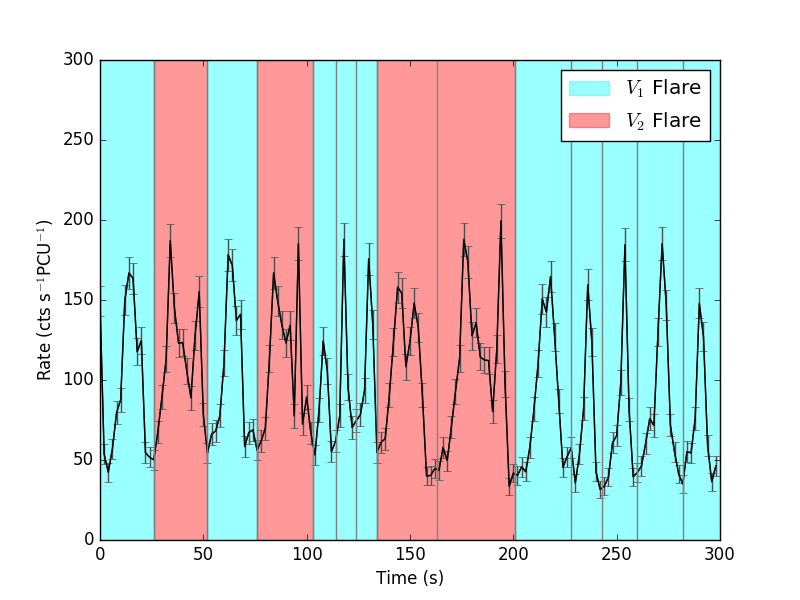}
    \captionsetup{singlelinecheck=off}
    \caption{A portion of the lightcurve of observation 96420-01-06-03, orbit 0, showing Type $V_1$ flares (highlighted in cyan) and Type $V_2$ flares (highlighted in red).}
   \label{fig:id_flares_V}
\end{figure}

\begin{figure}
    \includegraphics[width=\columnwidth, trim = 0mm 0mm 0mm 0mm]{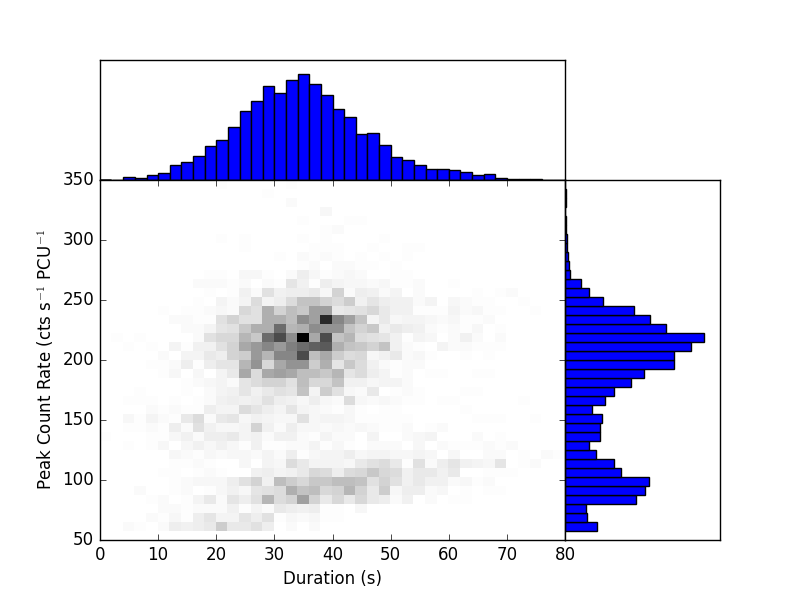}
    \captionsetup{singlelinecheck=off}
    \caption{Every flare in all observations identified as Class V, plotted in a two-dimensional histogram of flare peak count rate against flare duration to show the two-population nature of these events.}
   \label{fig:two_popV}
\end{figure}

 In Figure \ref{fig:id_flares_V} we show that the flaring in this class is more complex than that seen in Classes III and IV.  Class V lightcurves consist of short strongly peaked symmetrical flares (hereafter Type $V_1$) and a longer more complex type of flare (hereafter Type $V_2$).  The Type $V_2$ flare consists of a fast rise to a local maximum in count rate, followed by a $\sim10$\,s period in which this count rate gradually reduces by $\sim50\%$ and then a much faster peak with a maximum count rate between 1 and 2 times that of the initial peak.  In both types of flare, we find that the increase in count rate corresponds with an increase in soft colour.  The two-population nature of flares in Class V can also clearly be seen in Figure \ref{fig:two_popV}, where we show a two-dimensional histogram of flare peak count rate against flare duration.
\par
\begin{figure}
    \includegraphics[width=\columnwidth, trim = 0mm 0mm 0mm 0mm]{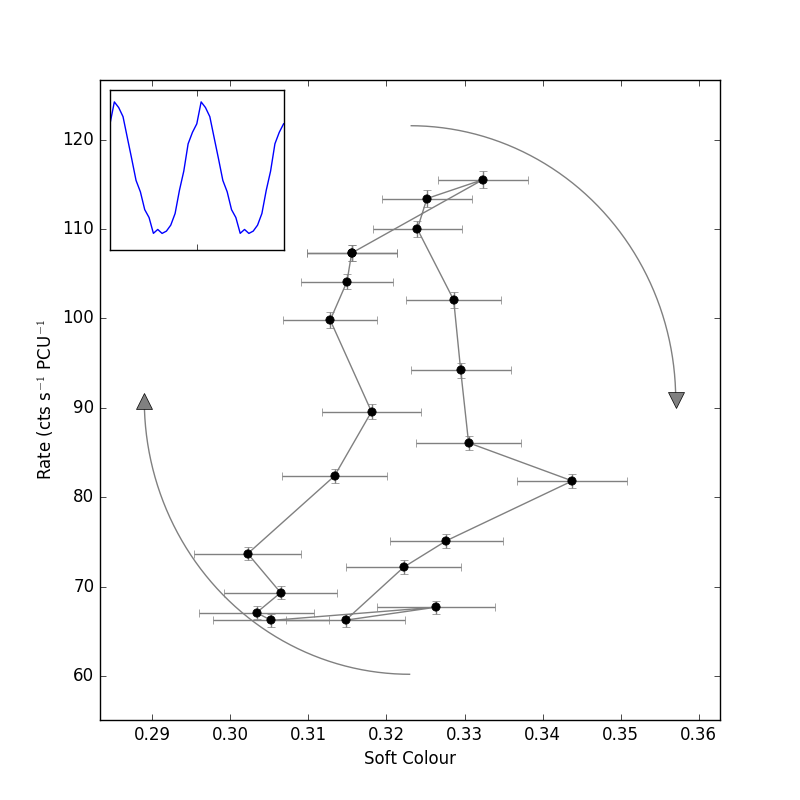}\\
    \includegraphics[width=0.5\columnwidth, trim = 0mm 0mm 0mm 0mm]{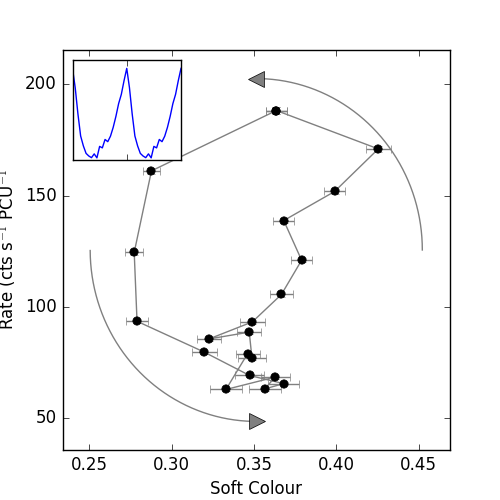}\includegraphics[width=0.5\columnwidth, trim = 0mm 0mm 0mm 0mm]{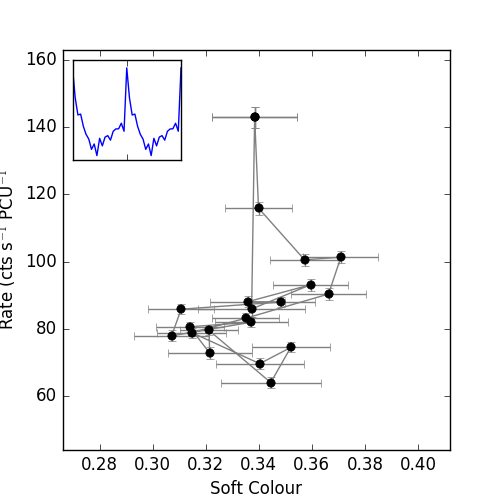}
    \captionsetup{singlelinecheck=off}
    \caption{\textit{Top}: The hardness-intensity diagram (HID$_1$) of a type $V_1$ flaring region in Class V observation 96420-01-07-00, orbit 0 showing a clockwise loop.  The data have been folded over a variable period found with the algorithm described in Appendix \ref{app:Flares}.  Inset is the folded lightcurve of the same data. \textit{Bottom Left}: The hardness-intensity diagram of Class V observation 96420-01-25-05 orbit 0, an example of an anticlockwise loop.  \textit{Bottom Right}: The hardness-intensity diagram of Class V observation 96420-01-25-06 orbit 0, in which we were unable to ascertain the presence of a loop.}
   \label{fig:LoopV}
\end{figure}

 We folded all individual Class V lightcurves, in each case cropping out regions of $V_2$ flaring.  We find clockwise hysteretic loops in the HID$_1$ of 30 out of 33 Class V observations, suggesting a lag in the aforementioned relation between count rate and soft colour.  In the upper panel Figure \ref{fig:LoopV} we present an example of one of these loops.  In one observation however, we found an anticlockwise loop in the HID$_1$ (shown in Figure \ref{fig:LoopV} lower-left panel).  We were unable to ascertain the presence of loops in the remaining 2 orbits; for the sake of completeness, we show one of these in the lower-right panel of Figure \ref{fig:LoopV}.

\begin{figure}
    \includegraphics[width=0.8\columnwidth, trim = 0.6cm 0 3.9cm 0]{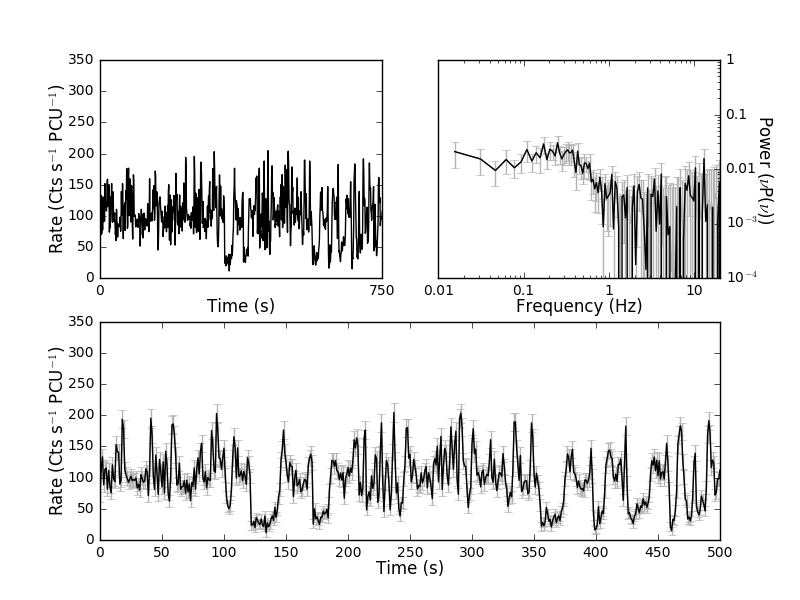}
    \captionsetup{singlelinecheck=off}
    \caption{Plots of the Class VI observation 96420-01-09-00, orbit 0.  \textit{Top-left}: 750\,s lightcurve binned on 2 seconds to show lightcurve evolution.  \textit{Top-right}: Fourier Power Density Spectrum.  \textit{Bottom}: Lightcurve binned on 1 second.}
   \label{fig:Lmulti}
\end{figure}

\subsubsection{Class VI -- Figure \ref{fig:Lmulti}}

\par The lightcurves of observations of this class show large dips in count rate; this can be seen in Figure \ref{fig:Lmulti} at, for example, $t\approx125$--$150$\,s .  These dips vary widely in duration, from $\sim5$ to $\sim50$ seconds, and the count rate in both $L_A$ and $L_B$ fall to a level consistent with background.  The dips' rise and fall times are fast, both lasting no longer than a second.  They do not appear to occur with any regular periodicity.
\par Aside from the dips, Class VI observations show other structures in their lightcurves.  Large fluctuations in count rate, by factors of $\lesssim3$, occur on timescales of $\sim1\mbox{--}5$ s; no periodicity in these oscillations could be found.  This behaviour is reflected in the PDS, which shows high-amplitude broad band noise below $\sim0.5$Hz with RMS-normalized power \citep{Belloni_RMSNorm} of up to $\sim1.1 $Hz$^{-1}$.  As can be seen in Figure \ref{fig:Lmulti}, this feature takes the form of a broad shoulder of noise which shows a either weak peak or no clear peak at all.  The $\sim5$Hz QPO seen in the PDS of other classes is not present in Class VI observations.
\par
\begin{figure}
    \includegraphics[width=\columnwidth, trim = 0mm 0mm 0mm 0mm]{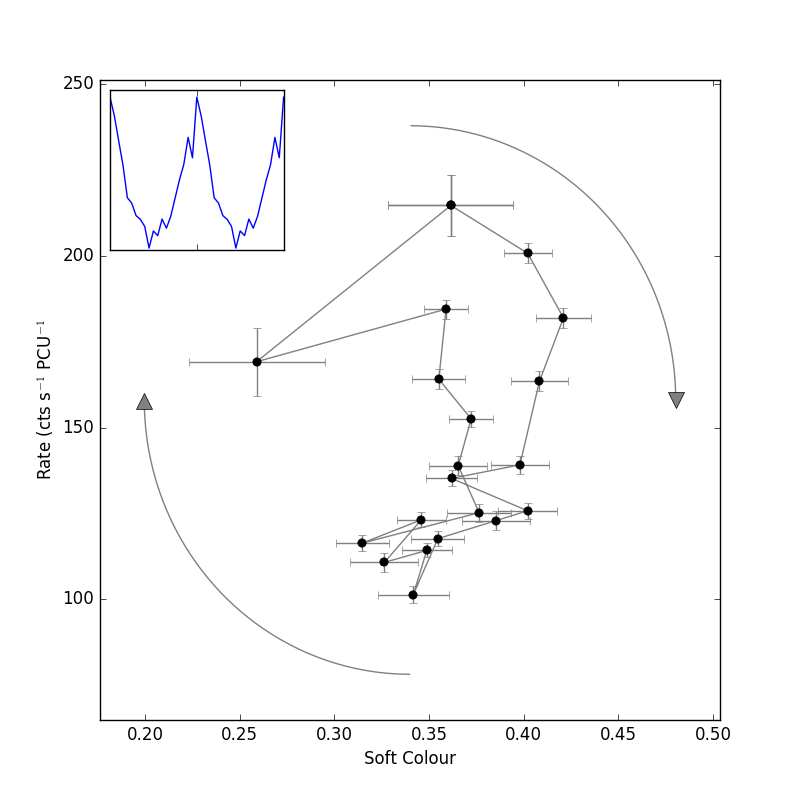}\\
    \includegraphics[width=0.5\columnwidth, trim = 0mm 0mm 0mm 0mm]{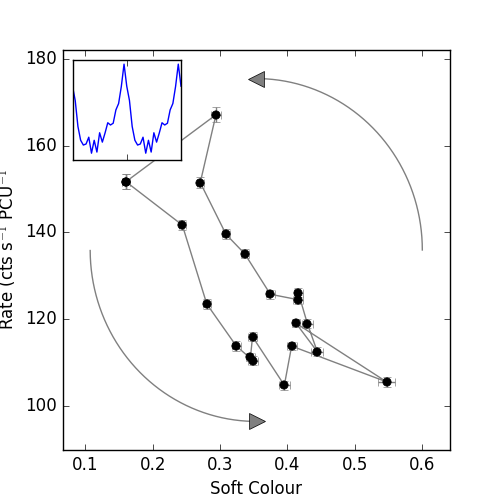}\includegraphics[width=0.5\columnwidth, trim = 0mm 0mm 0mm 0mm]{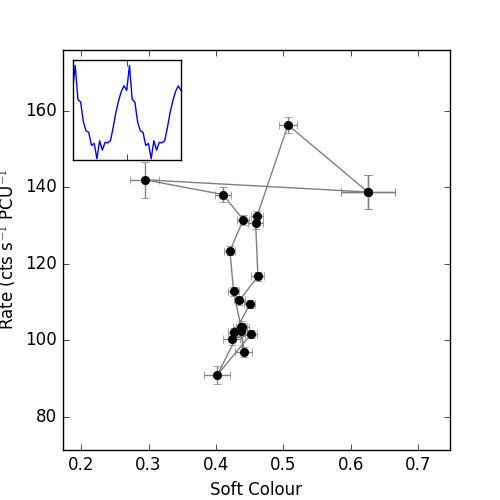}
    \captionsetup{singlelinecheck=off}
    \caption{\textit{Top}: The hardness-intensity diagram (HID$_1$) of the Class VI observation 96420-01-30-03, orbit 0 showing a clockwise loop.  The data have been folded over a variable period found with the algorithm described in Appendix \ref{app:Flares}.  Inset is the folded lightcurve of the same data. \textit{Bottom Left}: The hardness-intensity diagram of Class VI observation 96420-01-30-04 orbit 0, an example of an anticlockwise loop.  \textit{Bottom Right}: The hardness-intensity diagram of Class VI observation 96420-01-09-03 orbit 0, in which we were unable to ascertain the presence of a loop.}
   \label{fig:LoopVI}
\end{figure}

 We attempted to fold all individual Class VI lightcurves, ignoring the sections of data corresponding to the large count rate dips described above.  In general, folding lightcurves belonging to this class is difficult; many orbits showed low-amplitude oscillations which were difficult to fold using our flare-finding algorithm (see Appendix \ref{app:Flares}), while many others only showed oscillatory behaviour for a small number of periods between each pair of dips.  As such, we only succesfully folded 23 of the 40 Class VI orbits.  Of these, 19 showed clockwise loops in the HID$_1$ (top panel, Figure \ref{fig:LoopVI}), 3 showed anticlockwise loops (bottom-left panel, Figure \ref{fig:LoopVI}).  In the remaining 1 observation, the data did not allow us to ascertain the presence of loops (bottom-right panel, Figure \ref{fig:LoopVI}).
\par Like in Class VI, we note that the clockwise loops in Class VI appear more complex than clockwise loops.  Again, the clockwise loop shown in Figure \ref{fig:LoopVI} appears to have a 2-lobe structure; this is repeated in all clockwise loops found in this class.

\begin{figure}
    \includegraphics[width=0.8\columnwidth, trim = 0.6cm 0 3.9cm 0]{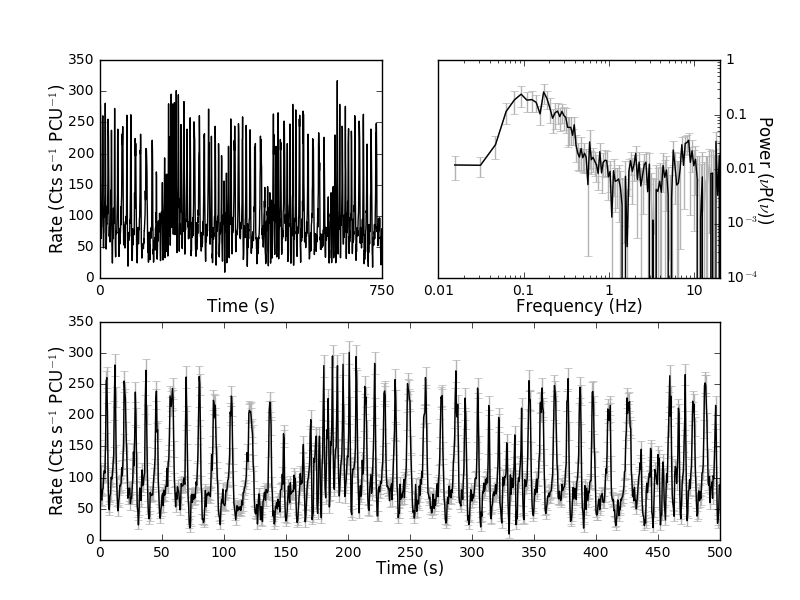}
    \captionsetup{singlelinecheck=off}
    \caption{Plots of the Class VII observation 96420-01-18-05, orbit 0.  \textit{Top-left}: 750\,s lightcurve binned on 2 seconds to show lightcurve evolution.  \textit{Top-right}: Fourier Power Density Spectrum.  \textit{Bottom}: Lightcurve binned on 0.5 seconds.}
   \label{fig:Nmulti}
\end{figure}

\subsubsection{Class VII -- Figure \ref{fig:Nmulti}}

\par
\begin{figure}
    \includegraphics[width=0.8\columnwidth, trim = 0.6cm 0 3.9cm 0]{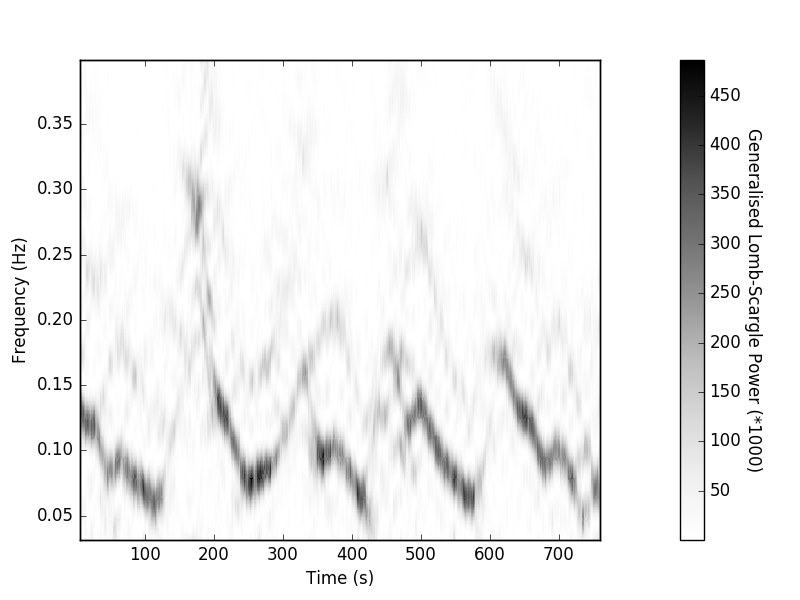}
    \captionsetup{singlelinecheck=off}
    \caption{A sliding window  Lomb-Scargle spectrogram of Class VII observation 96420-01-18-05, showing power density spectra from an overlapping 32\,s window moved 1\,s at a time.  The peak frequency of this low frequency QPO itself appears to oscillate with a frequency of $\sim5$mHz.}
   \label{fig:spect}
\end{figure}

 Class VII shows high-amplitude flaring behaviour with a peak-to-peak recurrence time of $6$--$12$\,s.  In Figure \ref{fig:spect} we show a dynamical Lomb-Scargle spectrogram of a Class VII observation, showing that the fast flaring behaviour has a frequency which moves substantially over time.  This in turn accounts for the large spread in the value of the flare peak-to-peak recurrence time.
\par In Figure \ref{fig:spect} we show that the peak frequency of the QPO also varies in a structured way.  We also suggest that the variabilitity of the frequency is itself a QPO with a period of $\sim150$.
\par At higher frequencies, the PDS shows a weak QPOs centred at $\sim8$Hz, with a $q$-values of $\sim2$.
\par We used our flare-finding algorithm (see Appendix \ref{app:Flares}) to perform variable-frequency folding of Class VII orbits.  We find clockwise loops in 9 out of 11 Class VII orbits.  In the remaining two observations, the oscillations were extremely fast.  As a result, the errors in the HID$_1$ of these too observations were too large to succesfully select peaks, and we are unable to confirm or reject the presence of loops.

\begin{figure}
    \includegraphics[width=0.8\columnwidth, trim = 0.6cm 0 3.9cm 0]{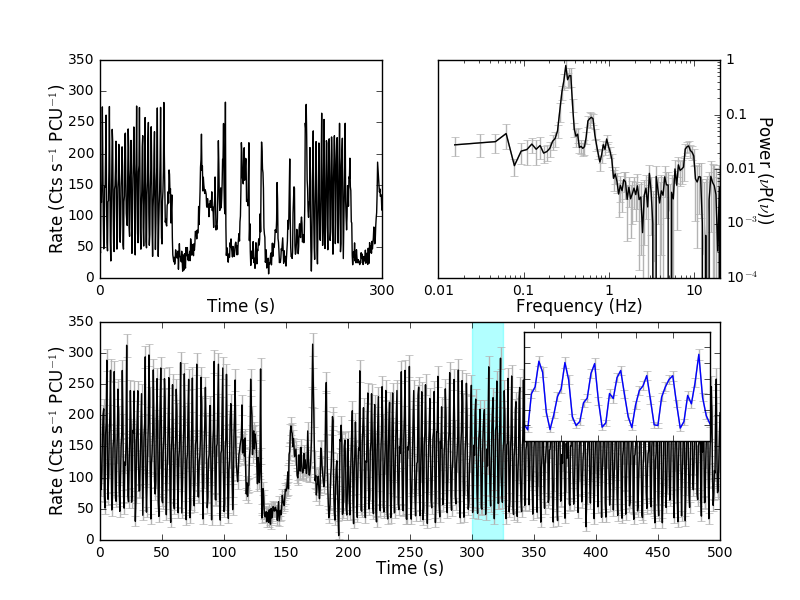}
    \captionsetup{singlelinecheck=off}
    \caption{Plots of the Class VIII observation 96420-01-19-03, orbit 0.  \textit{Top-left}: 300\,s lightcurve binned on 2 seconds to show lightcurve evolution.  \textit{Top-right}: Fourier Power Density Spectrum.  \textit{Bottom}: Lightcurve binned on 0.5 seconds.  Inset is a zoom of the 25\,s portion of the lightcurve highlighted in cyan, to show the second-scale structure in the lightcurve.}
   \label{fig:Omulti}
\end{figure}

\subsubsection{Class VIII -- Figure \ref{fig:Omulti}}

\par The lightcurve of this variability class shows the dipping behaviour seen in Class VI, as can be seen in Figure \ref{fig:Omulti} at $t\approx125$--$150$\,s.  The dips are less frequent than in Class VI.  The behaviour outside of the dips is dominated by highly structured high-amplitude oscillations consisting of flares with a peak to peak separation of $3.4\pm1.0$\,s.  The PDS shows this behaviour as a very significant ($q$-value > 20) QPO; two harmonics of this QPO are also visible.  The PDS also shows a strong ($q$-value = 4.7) QPO at $\sim9$Hz.
\par We attempted to fold Class VIII lightcurves, ignoring the portions of data corresponding to dips, using our flare-finding algorithm.  The high frequency of the dominant oscillation in Class VIII resulted in large errors in the peak times of individual flares, which translated to large errors in all HID$_1$s; however, we were able to ascertain the presence in loops in 8 out of 16 orbits.  All 8 of these loops are clockwise.

\begin{figure}
    \includegraphics[width=0.8\columnwidth, trim = 0.6cm 0 3.9cm 0]{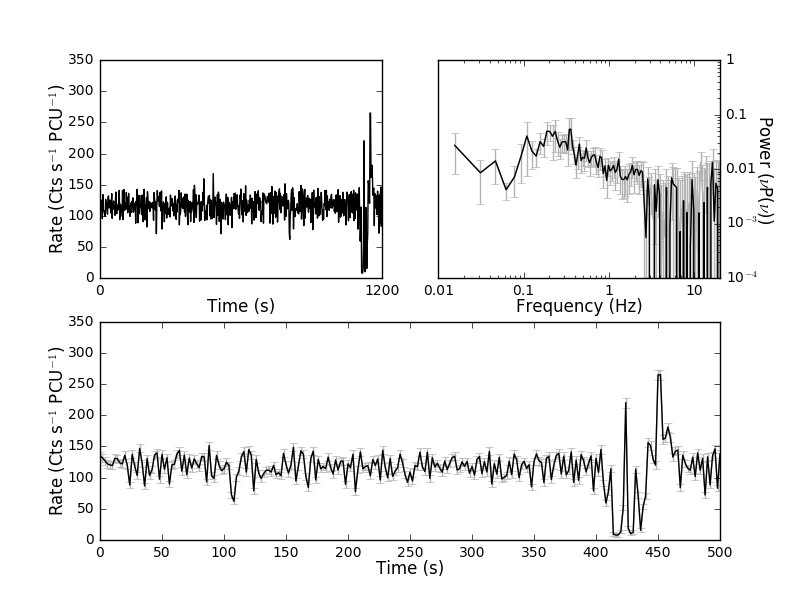}
    \captionsetup{singlelinecheck=off}
    \caption{Plots of the Class IX observation 96420-01-35-02, orbit 1.  \textit{Top-left}: 1200\,s lightcurve binned on 2 seconds to show lightcurve evolution.  \textit{Top-right}: Fourier Power Density Spectrum.  \textit{Bottom}: Lightcurve binned on 2 seconds.}
   \label{fig:Qmulti}
\end{figure}

\subsubsection{Class IX -- Figure \ref{fig:Qmulti}}
\label{sec:ClassIX}

\par The 1\,s lightcurve of a Class IX observation is superficially similar to the lightcurve of a Class I observation, with little obvious structured variability at timescales larger than 2 s; however, large count rate dips like those seen in Classes VI and VIII (e.g. the feature at $t\approx410$\,s in the lightcurve of Figure \ref{fig:Qmulti}) are very occasionally observed.  These dips may in turn be coupled to short second-scale flares in which count rate briefly increases by a factor of 2--3.
\par Outside of these dips and flares, the lightcurve of a Class IX observation is indistinguishable from the lightcurve of a Class I or Class II observation.  However, in Figure \ref{fig:IIIisHarder}, we show that Class IX occupies a very different part of the global $H_{A2}$/$H_{A1}$ colour-colour diagram.  Class IX observations show a significantly larger $H_{A2}$ than Class I and II observations, but a significantly lower $H_{A1}$.
\par The PDS reveals significant broad band noise peaked at $\sim$0.3 Hz, and the $\sim5$Hz QPO seen in other classes is absent.  \citet{Altamirano_HFQPO} discovered high frequency ($\sim66$Hz) QPOs in observations corresponding to this variability class.

\subsection{Swift}

\par Observations with \textit{Swift} took place throughout the 2011-2013 outburst of IGR J17091-3624.  Between MJDs 55622 and 55880, 17 \textit{Swift/XRT} were at least partly simultaneous with an \rxte\ observation, corresponding to at least one observation of all 9 classes.  In each case, the \textit{Swift} and \rxte\ lightcurves were similar.  The remainder of the \textit{Swift/XRT} observations during this time were also consistent with belonging to one of our nine classes.  Given that the \rxte\ data have higher count rate and time resolution, we do not further discuss the \textit{Swift} observations taken before MJD 55880.  A more detailed comparison of \rxte\ and \textit{Swift} data is beyond the scope of this paper.
\par Between MJD 55952 and 56445, \textit{Swift} observations showed IGR J17091-3624 decreasing in flux.  For all observations longer than 500 s, we rebinned the lightcurves to 10\,s and calculated the RMS.  We find the lower and upper quartiles of the fractional RMS in these measurements to be 18.3\% and 21.7\% respectively.  \textit{INTEGRAL} observations taken as part of a scan programme of the Galactic Plane \citep{Fiocchi_PlaneScan} and reported by \citet{Drave_Return} suggest that IGR J17091-3624 returned to the hard state between MJDs 55952 and 55989.  Therefore these observations sample IGR J17091-3624 the hard state.

\subsection{INTEGRAL}

\par The results of the \textit{INTEGRAL}/IBIS analysis are presented in Table \ref{tab:IBIS_results}. We see clear detections of IGR J17091-3624 in all energy bands during the hardest period (MJD 55575--55625) of the 2011--2013 outburst. Conversion from detected counts to flux was achieved using an \textit{INTEGRAL}/IBIS observation of the Crab taken between MJD 57305.334 and 57305.894. Conversion from Crab units to standard flux units was obtained by conversion factors listed in \citet{Bird_Survey} and \citet{Bazzano_Survey}.

\begin{table*}
\begin{tabular}{cccccc}
\hline
\hline
Energy 		& Intensity 		& Significance 	& Exposure 	& Flux 				& Flux					\\
(keV)		& (cts/s)			& $\sigma$		& (ks)		& (mCrab) 			& (10$^{-10}$ergs~s$^{-1}$~cm$^{-2}$) 	\\
\hline
20--40		& 12.39$\pm$0.05	& 247			& 115		& 93.5$\pm$0.38		& 7.08$\pm$0.03			\\
40--100		& 7.06$\pm$0.05		& 157			& 163		& 83.5$\pm$0.60		& 7.87$\pm$0.06			\\
100--150	& 1.05$\pm$0.03		& 40			& 173		& 66.9$\pm$1.91		& 2.14$\pm$0.06			\\
150--300	& 0.23$\pm$0.03		& 7.6			& 179		& 46.6$\pm$5.96		& 2.24$\pm$0.29			\\	
\hline
\hline
\end{tabular}
\caption{Results from the IBIS/ISGRI analysis of the 2011--2013 Outburst of IGR J17091. The 20--40\,keV flux is given in units of mCrab and (10$^{-11}$\ergf ). Conversion between counts and mCrab was obtained using an observation of the Crab taken during Revolution 1597 between MJD 57305.334 and 57305.894 and the conversion factors of \citet{Bird_Survey} and \citet{Bazzano_Survey}.}
\label{tab:IBIS_results}
\end{table*}

\par
\begin{figure}
    \includegraphics[width=0.7\columnwidth, trim = 0.6cm 0 3.9cm 0]{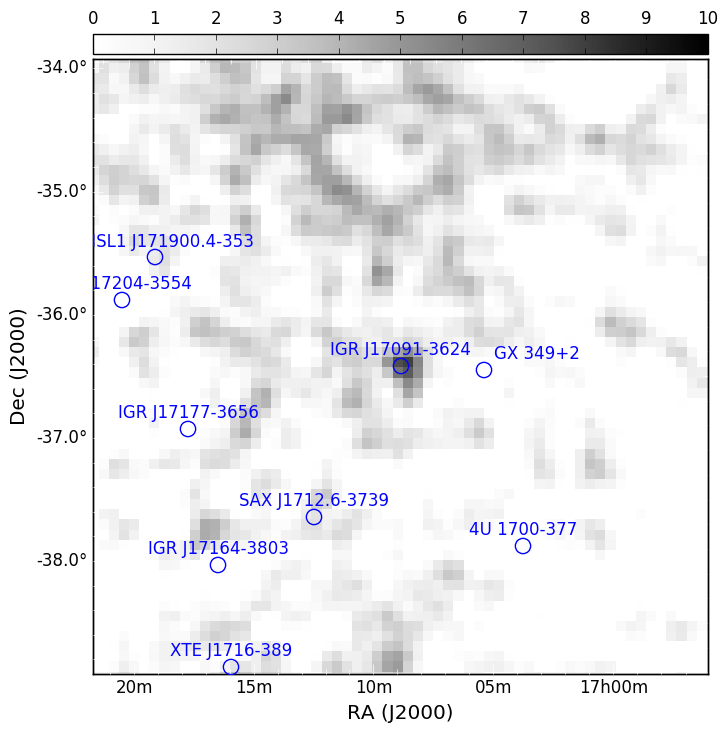}
    \captionsetup{singlelinecheck=off}
    \caption{\textit{INTEGRAL}/ISGRI 150--300\,keV significance map of a $2^\circ$ region centred on the position of IGR J17091-3624, showing the first significant detection of this source above 150\,keV.  The detection significance is 7.6 $\sigma$.}
   \label{fig:sigmap}
\end{figure}

 Comparing these results with those of \citet{Bazzano_Survey}, we see that IGR J17091 is detected for the first time above 150\,keV with a detection significance of 7.6\,$\sigma$, corresponding to a flux of $2.24\pm0.29\times10^{-10}$\ergf\ (Figure \ref{fig:sigmap}).

\subsection{Chandra}

\par
\begin{figure}
    \includegraphics[width=0.8\columnwidth, trim = 0.6cm 0 3.9cm 0]{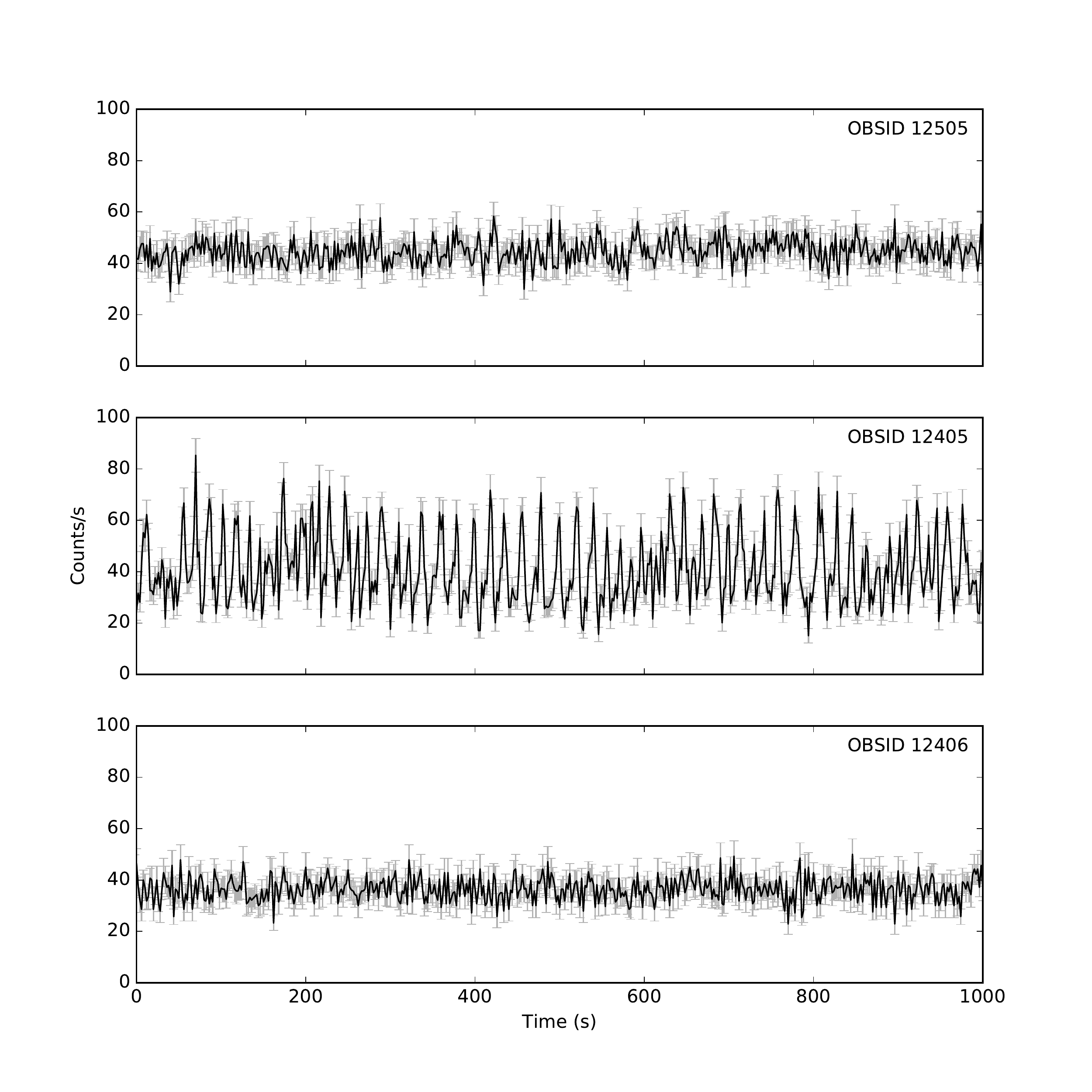}
    \captionsetup{singlelinecheck=off}
    \caption{1 ks segments of lightcurves taken from \textit{Chandra} observations 12505, 12405 and 12406, showing Class I, Class VII and Class IX variability respectively.  The lightcurve presented for observation 12505 is for the energy range 0.06-10\,keV, while the other two lightcurves are for the energy range 0.5-10\,keV.  All three lightcurves are binned to 0.5\,s.}
   \label{fig:Cha_lc}
\end{figure}

 In Figure \ref{fig:Cha_lc}, we present lightcurves from the three \textit{Chandra} observations considered in this paper (see also Table \ref{tab:Chandra} for details of these observations).

\par Observation 12505 was performed within 24 hours of \rxte\ observation 96420-01-02-01, which showed Class I variability.  No structured variability is seen in the lightcurve of OBSID 12505 (Figure \ref{fig:Cha_lc}, upper panel), which is consistent with Class I.  Note that we consider the energy range 0.06-10\,keV for this observation but 0.5-10\,keV for observations 12405 and 12406.
\par Observation 12405 was performed within 24 hours of \rxte\ observation 96420-01-23-03, which showed Class V variability.  The two observations were not simultaneous; OBSID 12405 began $\sim8.4$ ks after OBSID 96420-01-2303 finished.  The lightcurve of \textit{Chandra} OBSID 12405 (shown in Figure \ref{fig:Cha_lc}, middle panel) shows a mean count rate of 41\,cts\,s$^{-1}$.  The lightcurve shows fast flaring behaviour (with a recurrence time on the order of 10s of seconds) in which the frequency changes widely on timescales of $\sim1000$\,s.  This observation strongly resembles a Class VII lightcurve, but with its characteristic timescales increased by a factor of $\sim4$.  This leads to the possibility that the low number of Class VII \rxte\ observations we identify is due to a selection effect; we would not have been able to see this observation's long-term Class VII-like behaviour if the observation had been shorter than $\sim2$ ks.
\par Observation 12406 was performed within 24 hours of \rxte\ observation 96420-01-32-06, which showed Class IX variability.  The lightcurve presented for \textit{Chandra} OBSID 12406 shows a mean count rate (36 cts s$^{-1}$), which is consistent with IGR J17091 being harder in this observation than in Observation 12505.  This, combined with the lack of variability seen in its lightcurve, suggests that Observation 12505 is consistent with Class IX.

\subsection{XMM-Newton}

\par
\begin{figure}
    \includegraphics[width=0.8\columnwidth, trim = 0.6cm 0 3.9cm 0]{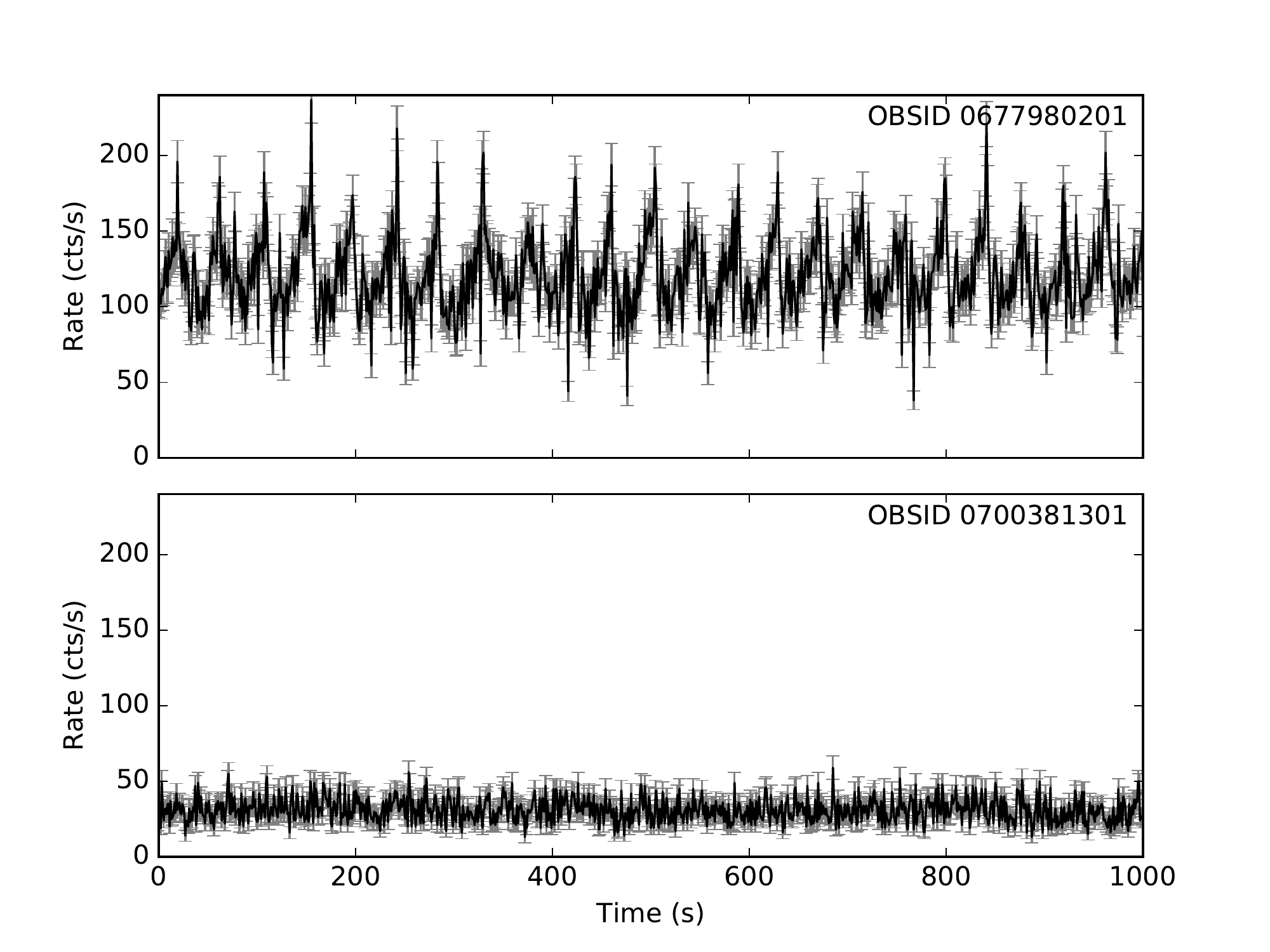}
    \captionsetup{singlelinecheck=off}
    \caption{Lightcurves of \textit{XMM-Newton} observations 0677980201 and 0700381301, showing Class IV variability and the hard state respectively.  Both lightcurves binned to 2\,s.  Data for observation 0677980201 is taken from \textit{EPIC-MOS2} and data for observation 0700381301 is taken from \textit{EPIC-pn}.}
   \label{fig:XMM}
\end{figure}

 In Figure \ref{fig:XMM} we show lightcurves from two \textit{XMM-Newton} observations.  The lightcurve of \textit{XMM-Newton} observation 0677980201, shown in the upper panel of Figure \ref{fig:XMM}, shows the regular flares characteristic of Class IV variability.  A simultaneous \rxte\ observation (OBSID 96420-01-05-000) also showed Class IV variability.

\par
\begin{figure}
    \includegraphics[width=0.9\columnwidth, trim = 0cm 0cm 0.5cm 1.0cm, clip]{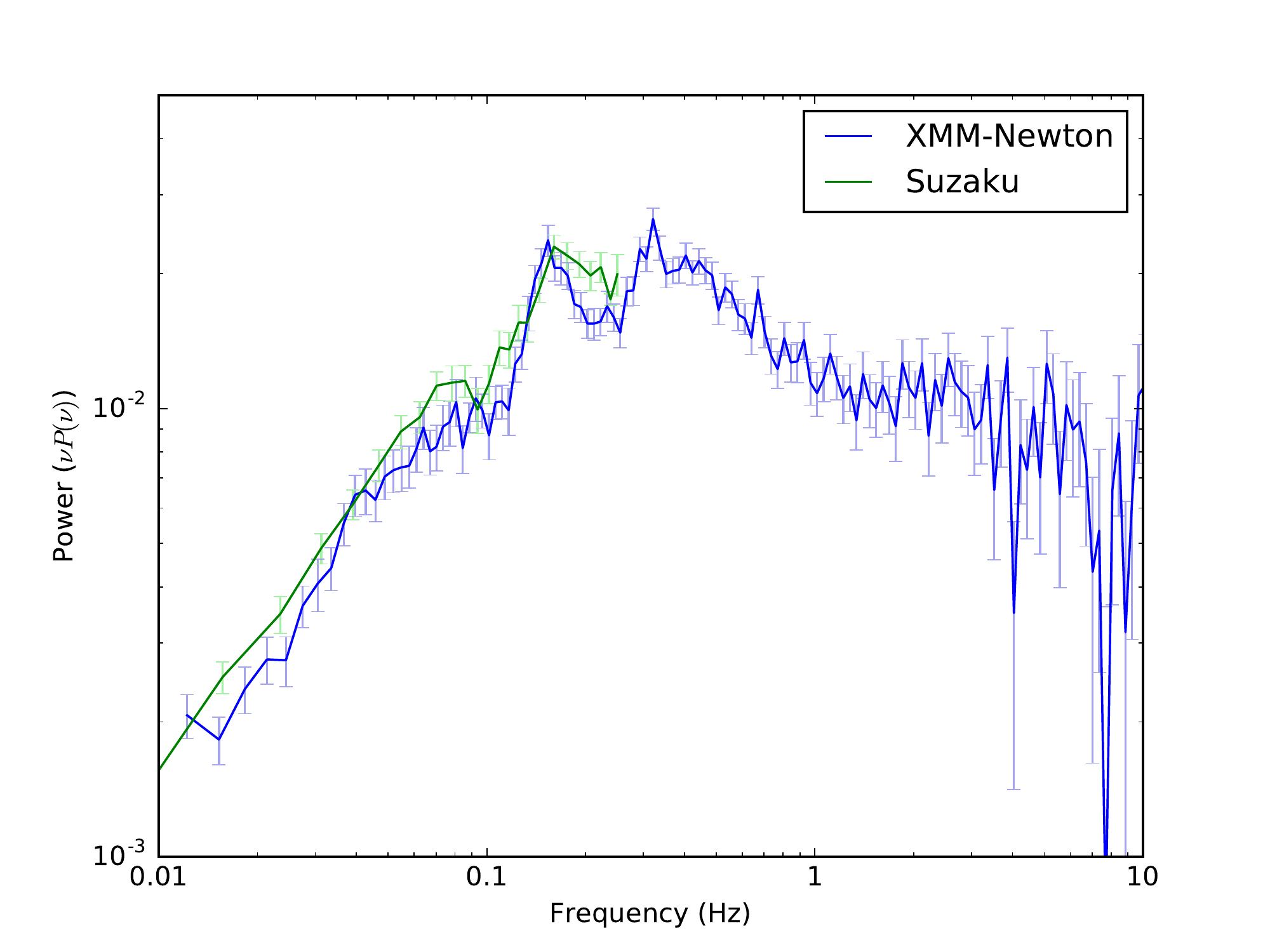}
    \captionsetup{singlelinecheck=off}
    \caption{$\nu P(\nu)$-normalised co-added power density spectra of \textit{XMM-Newton} observation 0700381301 and \textit{Suzaku} observation 407037010.  Both observations were taken simultaneously on September 29 2012 (MJD 56199).  We sample observation 0700381301 up to a frequency of 10\,Hz, while the 2\,s time resolution of observation 407037010 results in a Nyquist frequency of 0.25\,Hz.}
   \label{fig:xmmqpo}
\end{figure}

 \textit{XMM-Newton} observation 070038130, shown in the lower panel of Figure \ref{fig:XMM}, was made after the end of \rxte\ observations IGR J17091-3624.  As such it cannot be compared with contemporaneous \rxte\ data.  The 5\,s binned lightcurve shows no apparent variability, but a Fourier PDS of the observation (shown in Figure \ref{fig:xmmqpo}) reveals a QPO centred at around $\sim0.15$Hz and a broad band noise component at lower frequencies.  \citet{Drave_Return} reported that IGR J17091 transited to the hard state in February 2012, seven months before this observation was taken.  As such, we find that observation 0677980201 samples the hard state in IGR J17091 and is thus beyond the scope of our set of variability classes.

\subsection{\textit{Suzaku}}

\par The two {\it Suzaku} observations of IGR J17091-3624 considered, OBSIDs 407037010 and 407037020, were performed during the 2nd and 3rd re-flares of the hard state phase of the 2011--2013 outburst.  OBSID 407037010 was taken simultaneously with \textit{XMM-Newton} observation 0700381301.  The XIS 0 count rates are 7.8 cts\,s$^{-1}$ and 2.5\,cts\,s$^{-1}$ respectively.
\par Neither lightcurve shows `heartbeats' or any other type of GRS 1915-like variability.  However, we find evidence of a low frequency QPO feature at $\sim$0.15 Hz in the OBSID 407037010; this QPO is also seen in \textit{XMM-Newton} observation 0700381301 (Figure \ref{fig:xmmqpo}).  The presence of a QPO below 1\,Hz and flat-topped power density spectrum confirm that IGR J17091 was in the hard state at this time.

\section{Discussion}

\par Using observations from \textit{XMM-Newton}, \rxte\ and \textit{Chandra}, we describe the complex variability seen in IGR J17091 as a set of nine variability `classes', labelled I to IX.  These classes are distinguished from each other by values of upper and lower quartile (i.e. 25\textsuperscript{th} and 75\textsuperscript{th} percentile) count rates, mean RMS, the presence of QPOs in Fourier PDS, the shape of flare and dip features in the lightcurve and the presence of loops in the 6--16/2--6 keV hardness-intensity diagram HID$_1$.  See Section \ref{sec:results} for a full description of these classes.
\par
\begin{figure}
    \includegraphics[width=\columnwidth, trim =0cm 0 0cm 0]{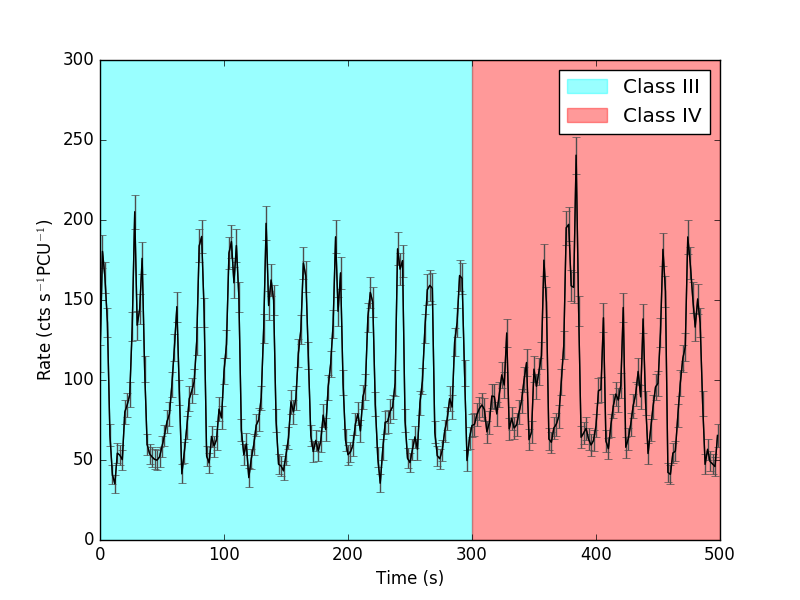}
    \captionsetup{singlelinecheck=off}
    \caption{A lightcurve of observation 96420-01-06-02, orbit 0, showing a transition in behaviour between Class IV (in cyan, see Section \ref{sec:classIV}) and Class V (in red, see Section \ref{sec:classV}).}
   \label{fig:HybridClasses}
\end{figure}

 The classification of some observations is clearer than others.  Some orbits were too short to definitively quantify the behaviour of the source, whereas some other orbits contain a transition between two classes.  An example lightcurve showing a transition from Class III to Class IV is presented in Figure \ref{fig:HybridClasses}.

\par Our set of classes is analogous to, but not based upon, the set of variability classes defined by \citealt{Belloni_GRS_MI} to describe the behaviour of the similarly complex LMXB GRS 1915.  This ensures that our set of classes is not biased by an \textit{a priori} assumption that the two objects are similar.  However if we do assume that wide range of variability seen in these two objects are driven by the same physical processes, a direct comparison between the variability classes in the two systems can further our understanding of the physics that drive these exotic objects.
\par We also use all 2011-2013 IGR J17091-3624 data from \rxte , \textit{XMM-Newton}, \textit{Chandra}, \textit{Swift}, \textit{INTEGRAL} and \textit{Suzaku} to analyse the long-term evolution of the 2011--2013 outburst.  This in turn corresponds to all available X-ray data taken during this outburst.

\subsection{Variability Classes: IGR J17091 vs. GRS 1915}

\par As observations of IGR J17091 and GRS 1915 suffer from different values of interstellar absorption $N_H$, we cannot directly compare the absolute colours of these two objects.  However, we can compare the evolution of colour both over time and as a function of count rate.  We therefore use these parameters, along with power spectra and lightcurve morphology, when comparing GRS 1915 with IGR J17091.
\par For seven of our classes, we were able to assign the closest matching class described by \citealt{Belloni_GRS_MI} for GRS 1915 (see Table \ref{tab:class_assign}).  We are unable to find analogues to our classes VII and VIII in observations of GRS 1915, and we suggest that these classes are unique to IGR J17091.

\begin{table}
\centering
\caption{The nine variability classes of IGR J17091-3624, showing the name of the closest corresponding variability class in GRS 1915+105.  The names of GRS 1915+105 classes are taken from \citet{Belloni_GRS_MI}, where more detailed descriptions can be found.  Eight additional classes of GRS 1915+105 have been described; we do not find analogies to these classes in IGR J17091-3624.}
\label{tab:class_assign}
\begin{tabular}{cc} 
\hline
\hline
IGR J17091-3624 Class & GRS 1915+105 Class\\
\hline
I&$\chi$\\
II&$\phi$\\
III&$\nu$\\
IV&$\rho$\\
V&$\mu$\\
VI&$\lambda$\\
VII&\textit{None}\\
VIII&\textit{None}\\
IX&$\gamma$\\
\hline
\hline
\end{tabular}
\end{table}

\par Below, we evaluate our mapping between GRS 1915 and IGR J17091 classes, and interpret the differences between each matched pair.

\subsubsection{Classes I and II -- Figures \ref{fig:Bmulti}, \ref{fig:Emulti}}

\label{sec:DisI}

\par Classes I and II both show low count rates and little structure in their lightcurves.  The two classes in GRS 1915 that also show this lightcurve behaviour are Class $\chi$\footnote{Note that, in GRS 1915+105, Class $\chi$ is further subdivided into four classes based on hard colour \citep{Belloni_GRS_MI,Pahari_Chi}.  As we cannot obtain hard colour for IGR J17091, we treat $\chi$ as a single variability class here.} and Class $\phi$.  \citealt{Belloni_GRS_MI} differentiate between Classes $\phi$ and $\chi$ based on the hard colour (corresponding to $C_2$), as Class $\chi$ has a significantly higher value for this colour than Class $\phi$.

\par Data from \rxte\  indicates that the transition from the hard state to the soft intermediate state between MJDs 55612 and 55615 \citep{Drave_Return}.  This was confirmed by a radio spectrum taken on MJD 55623 which was consistent with an observation of discrete ejecta \citep{Rodriguez_D}.  This observation of discrete ejecta at the transition between the hard state and the intermediate state has been reported in other LMXBS (e.g. XTE J1550-564, \citealp{Rodriguez_XTE}), and has also been associated with transitions to the $\chi$ Class in GRS 1915 (\citealp{Rodriguez_Ejection}, see also review by \citealp{Fender_Jets}).

\par Using Fourier PDS, we conclude that Class I is analogous to Class $\chi$ in GRS 1915, while Class II is analogous to Class $\phi$.  In Class $\chi$ observations of GRS 1915, broad band noise between $\sim1-10$Hz and a QPO at around 5Hz are seen in the PDS.  We find that both of these are present in Class I observations of IGR J17091.  On the other hand, we find that Class $\phi$ observations of GRS 1915 do not show this broad band noise, and show either a weak ($q$-value $\lesssim 3$) QPO at $\sim5$Hz or no QPO at all.  We find that the weak QPO and lack of broad band noise are also seen in the PDS of Class II observations.

\subsubsection{Classes III and IV -- Figures \ref{fig:Gmulti}, \ref{fig:Jmulti}}

\par
\begin{figure}
    \includegraphics[width=\columnwidth, trim = 0mm 0mm 0mm 0mm]{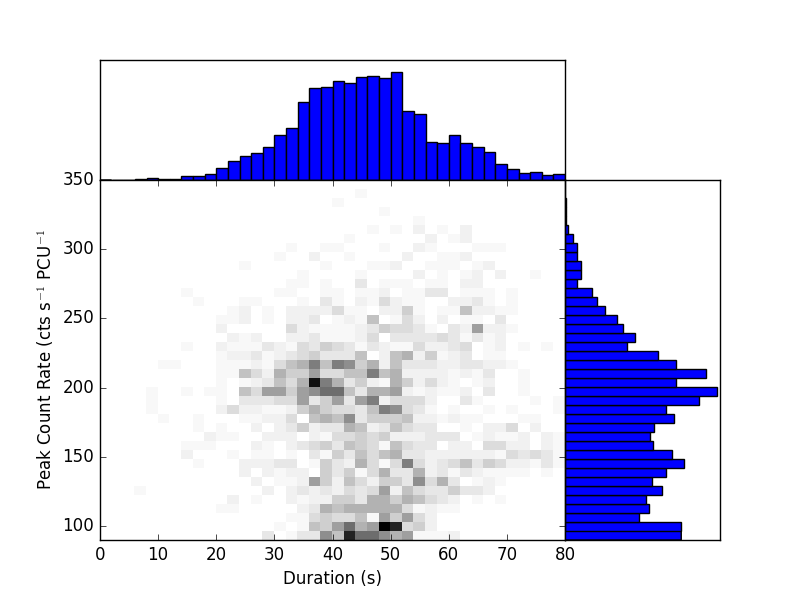}
    \captionsetup{singlelinecheck=off}
    \caption{Every flare in all observations identified as Class III or Class IV, plotted in a two-dimensional histogram of flare peak count rate against flare duration to show the two-population nature of these events.  Flares belonging to Class IV occupy the distribution at higher peak rate and lower duration, whereas flares belonging to Class III occupy the distribution at lower peak rate and higher duration.}
   \label{fig:III_IV_burst}
\end{figure}

 Classes III and IV both show highly regular flaring activity in their lightcurves, but they differ in terms of timescale and pulse profile.  As can be seen in lightcurves in Figure \ref{fig:Jmulti}, flares in Class IV occur every $\sim32$\,s and are nearly identical to each other in shape.  On the other hand, as can be seen in Figure \ref{fig:Gmulti}, flares in Class III occur every $\sim61$\,s and may or may not end in a much faster sharp peak which is never seen in Class IV.  In Figure \ref{fig:III_IV_burst} we show a two-dimensional histogram of flare peak count rate against flare duration, showing all flares in all observations classified as Class III or Class IV.  In this figure, we can see that flares tend to group in one of two regions in count rate-duration space; a region between $\sim90\mbox{--}110$ \spcu and $\sim35\mbox{--}55$\,s, corresponding to flares seen in Class III, and a region between $\sim150\mbox{--}250$ \spcu and $\sim20\mbox{--}55$\,s, corresponding to flares seen in Class IV.  From this plot, we conclude that the flares seen in Class III exist in a different population to the flares seen in Class IV.
\par The GRS 1915 classes that show behaviour most similar to these are $\rho$ and $\nu$; both produce similar structures in their lightcurve, but Class $\nu$ is differentiated from Class $\rho$ by the presence of a secondary count rate peak which occurs $\sim5$\,s after the primary \citep{Belloni_GRS_MI}.

\par
\begin{figure}
    \includegraphics[width=\columnwidth, trim = 0mm 0mm 0mm 0mm]{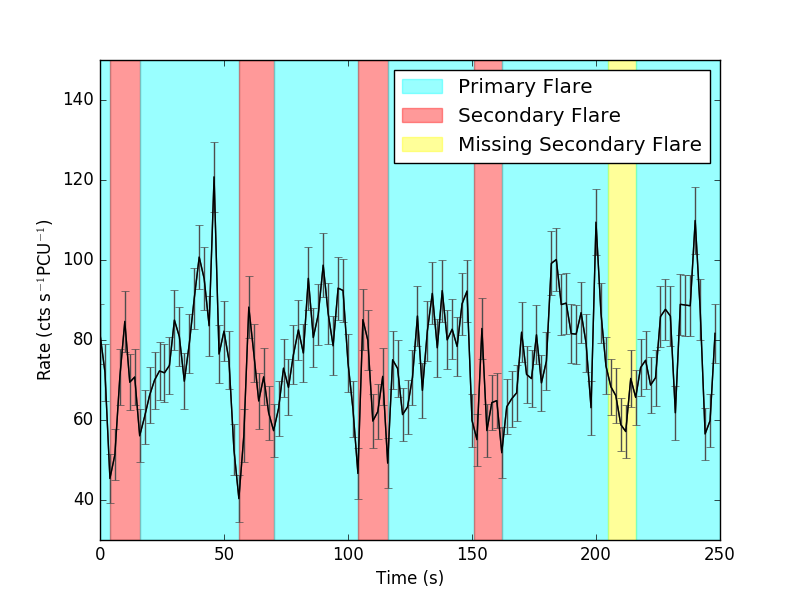}
    \captionsetup{singlelinecheck=off}
    \caption{Lightcurve from Class III observation 96420-01-10-01 of IGR J17091-3624, with pairs of primary and secondary count rate spikes highlighted in cyan and red respectively.  The yellow region highlights a primary count rate spike that did not produce a secondary.}
   \label{fig:III_IV_spike}
\end{figure}

 The secondary peak is present in most Class III observations and some Class IV observations (Figure \ref{fig:III_IV_spike}), suggesting that both classes consist of a mix of $\rho$-like and $\nu$-like observations.  However, the poor statistics sometimes make the presence of this secondary peak difficult to detect.  As such, we do not use the presence or absence of this peak as a criterion when assigning classes.  Instead we choose to separate Classes III and IV based on the larger-scale structure in their lightcurves (see Section \ref{sec:classIV}).  Due to the aforementioned difference in burst populations between the two classes, we suggest that classes III and IV do represent two distinct classes rather than a single class with a period that drifts over time.  We suggest that Classes $\rho$ and $\nu$ in GRS 1915 could also be re-partitioned in this way.

\par However, HID$_1$ loops are found to generally execute in an anticlockwise direction in Classes III and IV (previously noted by e.g. \citealp{Altamirano_IGR_FH}); the opposite direction to the clockwise loops in Classes $\rho$ and $\nu$ reported by e.g. \citealp{Belloni_GRS_MI} and repeated by us using the same method we apply to data from IGR J17091-3624 (see Section \ref{sec:dex}).  This suggests that Classes III and IV could be generated by a different physical mechanism to Classes $\rho$ and $\nu$.  Alternatively, Classes III and IV could be generated by the same mechanism as $\rho$ and $\nu$ if some other unknown process was able to alter the spectral evolution of flares in these classes.

\subsubsection{Class V -- Figure \ref{fig:Kmulti}}

\par The lightcurve of a Class V observation appears similar to that of a Class $\mu$ observation of GRS 1915, as both are characterised by rapid $\rho$-like flares which occur less regularly than in Class $\rho$.  In addition to this, flares in Class $\mu$ fall into two clear populations, as do the flares in Class V.  However, significant differences exist between Class V and Class $\mu$.  Class $\mu$ observations are characterised by long ($\sim100$ s) excursions to plateaus of high count rate, a behaviour which is not seen in any Class V observation thus far.
\par We note that the HID$_1$ in Class V observations displays a loop in the clockwise direction; the opposite direction to the looping seen in Classes III and IV but the same direction seen in Class $\mu$.
\par Regarding the two-population nature of flares seen in this class (see Section \ref{sec:classV}), we suggest that V$_2$ flares may simply be two V$_1$ flares that occur close together in time, such that the second flare starts during the decay of the first flare.  This would result in an apparent two-peaked flare structure, as we see in type V$_2$ flares.  This interpretation also accounts for the bimodal distribution of flare duarations shown in the 2D histogram of Figure \ref{fig:two_popV}, as this could be caused by the misinterpretation of two-flare V$_2$ events as a single event.  This also accounts for the Gaussian distribution of peak flare intensities seen in Figure \ref{fig:two_popV}), as the constituents of each V$_2$ event would be from the same population as V$_1$ flares.

\subsubsection{Class VI -- Figure \ref{fig:Lmulti}}

\par Class VI is dominated by long flaring periods which separate periods of low count rate, as can be seen in the lightcurve presented in Figure \ref{fig:Lmulti}.  Similar behaviour is seen in the lightcurves of observations of GRS 1915 belonging to Classes $\lambda$ and $\omega$ \citep{KleinWolt_OmegaClass}.  However, the long count rate `dips' are far less regular in Class VI than in Classes $\lambda$ and $\omega$, and we also note long periods of medium count rate during which neither flares nor dips occur.  This variability class is noted by \citet{Pahari_IGRClasses} who suggest that this class is unique to IGR J17091\footnote{\citet{Pahari_IGRClasses} refers to Class VI as Class C2.}.  However, \citet{Pahari_ClassVI} show that, in a plot of burst decay time against burst rise time, Classes VI and $\lambda$ fall in a straight line, suggesting a similar physical origin for both.
\par While it is cetainly true that Class VI is not a perfect analogue of either Class $\lambda$ or Class $\omega$, Class VI only differs noticeably from Class $\lambda$ during the extended low-variability portions of its lightcurves.  As such, we associate Class VI with Class $\lambda$.

\subsubsection{Class VII -- Figure \ref{fig:Nmulti}}

\par We are unable to find an analogue of Class VII in observations of GRS 1915.  This class, and its apparent uniqueness, have previously been noted by \citealp{Pahari_IGRClasses}\footnote{\citet{Pahari_IGRClasses} refers to Class VII as Class C1.}.  \citeauthor{Pahari_IGRClasses} found  that the $C_2$ hard colour in this class increases during count rate dips and decreases during count rate peaks.  Here we reproduced the results of \citeauthor{Pahari_IGRClasses} and found that the anti-correlation between hard-colour and intensity is not physical, but due to the definition of $C_2$: the count rate in band $L_C$ is approximately constant and consistent with background, and therefore $C_2=L_C/L_A \propto L_A^{-1}$, which will naturally anticorrelate with intensity.
\par Although a correlation between QPO frequency and count rate has been noted in the $\sim5$Hz QPO seen in GRS 1915 (e.g. \citealp{Markwardt_FluxFreqGRS,Vignarca_FluxFreqGRS}), this QPO is also seen in Class VII observations at the same time as the $\sim0.1$Hz QPO.  As such, the flux-frequency relationship in the very low frequency ($\sim0.1$Hz) QPO in Class VII is apparently unique amongst the classes of both IGR J17091 and GRS 1915.

\subsubsection{Class VIII -- Figure \ref{fig:Omulti}}

\par We are unable to find an analogue of Class VIII in observations of GRS 1915.  When it is flaring, the lightcurve waveform is similar to that seen in Class $\rho$, with rapid regular spikes in count rate.  The lightcurve also shows irregular dips in count rate similar to those seen in Class VI and in Class $\lambda$ in GRS 1915.
\par However, the amplitude of the flares in Class VIII is much larger, and the frequency much higher, than in Classes VI or $\lambda$.  The amplitude of the flares in Class VIII can approach $\sim350$cts s$^{-1}$ PCU$^{-1}$, while the flare separation time of 4--5\,s makes Class VIII the fastest flaring activity seen in any class of IGR J17091 or GRS 1915.  As such, we consider this variability class distinct from both Class VI and Class $\lambda$. 

\subsubsection{Class IX - Figure \ref{fig:Qmulti}}

\label{sec:DisIX}

\par Class IX is defined by long periods of high amplitude but unstructured variability (with a broad peaked noise component in the Fourier spectrum peaked at $\sim$0.3 Hz) punctuated with infrequent irregular short-duration `spikes' in which the count rate increases by a factor of $\sim2$--$3$.  A similarity between this Class and Class $\gamma$ in GRS 1915 has been previously noted by \citet{Altamirano_HFQPO}.  However, the irregular spikes seen in some Class IX lightcurves are not reproduced in Class $\gamma$ lightcurves of GRS 1915.

\subsection{General Comparison with GRS 1915+105}

\par Overall, variability in IGR J17091 tends to be faster than structurally similar variability in GRS 1915, as can be noted in Classes III and IV compared to Classes $\rho$ and $\nu$ (see also \citealp{Altamirano_IGR_FH}).  Additionally, IGR J17091 also displays highly structured variability unlike anything yet seen in GRS 1915, with classes VII and VIII in particular showing very fine detail in their lightcurves.
\par In total we find 2 variability classes which are seen in IGR J17091 but not in GRS 1915, compared with 8 that are seen in GRS 1915 but not in IGR J17091.  As relatively little data exists on GRS 1915-like variability in IGR J17091, the presence of classes in GRS 1915 that are not seen in IGR J17091 could simply be an observational effect.  It is unknown how long each variability class lasts for and, as such, additional variability classes could have occurred entirely while IGR J17091 was not being observed (however, see \citealp{Huppenkothen_ML} for a study on GRS1915 based on more than 16 years of data).  However, GRS 1915 has displayed variability classes consistently since its discovery in 1992, implying that the two classes seen only in IGR J17091 are either completely absent in GRS 1915 or that they occur with a much lower probability.  In either case, this implies physical differences between methods of generating GRS 1915-like variability in the two objects.  
\par As noted in sections \ref{sec:DisI} to \ref{sec:DisIX}, variability classes seen in both IGR J17091 and GRS 1915 show differences in the different objects.  In particular, we note the presence of irregular flares in Class IX which are not seen in the analogous Class $\gamma$.  If these classes are indeed generated by the same processes in both objects, the differences between them must represent physical differences between the objects themselves.
\par It has previously been noted that, while the hardness ratios in IGR J17091 and GRS 1915 during $\rho$-like classes are different, the fractional hardening between the dip and peak of each flare is consistent with being the same in both objects \citep{Capitanio_peculiar}.  This suggests that the same physical process is behind the `heartbeats' seen in both objects.
\par We note the presence of hysteretic HID$_1$ loops in some classes of both objects.  Although these loops are always clockwise in GRS 1915, they can be executed in either direction in IGR J17091.  Classes in IGR J17091 that show loops all have a preferred loop direction: anticlockwise in Classes III and IV and clockwise in classes V, VI, VII and VIII.  In cases where the loop direction was opposite to that expected for a given class, loop detections were generally only marginally significant.  In particular, we note that Classes IV and V tend to show loops in opposite directions, despite the similarities between their lightcurves and the $\rho$, $\nu$ and $\mu$ classes in GRS 1915.   The fact that IGR J17091 can show HID$_1$ loops in both directions suggests that an increase in soft emission can either precede or lag a correlated increase in hard emission from IGR J17091.  Whether soft emission precedes or lags hard emission is in turn is dependent on the variability class.
\par There are also non-trivial similarities between variability in the two objects.  We note the presence of a $\sim5$Hz QPO in many of the classes seen in IGR J17091, and this same 5Hz QPO is seen in lightcurves of GRS 1915.  Similarly \citet{Altamirano_HFQPO} reported the discovery of a 66Hz QPO in IGR J17091; a very similar frequency to the 67Hz QPO observed in GRS 1915 \citep{Morgan_QPO}.  It is not clear why these QPOs would exist at roughly the same frequencies in both objects when other variability in IGR J17091 tend to be faster.

\subsection{Comparison with the Rapid Burster}

\par In 2015, \citet{Bagnoli_RB} discovered the existence of two GRS 1915-like variability classes in the neutron star binary MXB 1730-335, also known as the `Rapid Burster'.  Specifically, \citet{Bagnoli_RB} note the presence of variability similar to Classes $\rho$ and $\theta$ in GRS 1915.
\par Class $\theta$-like variability, seen in \rxte\ observation 92026-01-20-02 of the Rapid Burster, is not closely matched by any of the classes we identify for IGR J17091.  However, the lightcurves of a Class $\theta$ observation feature large dips in count rate similar to those seen in Classes VI and VIII in IGR J17091.
\par Conversely, Class $\rho$-like variability is seen in all three objects.  \citet{Bagnoli_RB} note that the variability of the $\rho$-like flaring is slower in the Rapid Burster than in either GRS 1915 or IGR J17091. It has previously been suggested that the maximum rate of flaring in LMXBs should be inversely proportional to the mass of the central object (e.g. \citealp{Belloni_Timescales,Frank_Timescales}).  In this case, the fact that variability is faster in IGR J17091 than in GRS 1915 could simply be due to a lower black hole mass in the former object \citep{Altamirano_IGR_FH}.  However if variability in the Rapid Burster is assumed to be physically analogous to variability in these two black hole objects, then we note that a correlation between central object mass and variability timescale no longer holds.

\subsection{Comparison with Altamirano \textit{et al.} 2011}

\label{sec:Alta}
\par \citet{Altamirano_IGR_FH} identify 5 GRS 1915 variability classes in a subset of observations from the 2011-2013 outburst of IGR J17091: six of these observations are presented in Table \ref{tab:me_Diego} along with the best-fit GRS 1915 class that we assign it in this paper (see also Table \ref{tab:class_assign}).

\begin{table}
\centering
\caption{The six OBSIDs explicitly classified in \citet{Altamirano_IGR_FH}.  We also present the GRS 1915 class with which we implicitly label each OBSID in this paper.}
\label{tab:me_Diego}
\begin{tabular}{ccc} 
\hline
\hline
OBSID & Altamirano \textit{et al.} Class & Court \textit{et al.} Class\\
&&(implied)\\
\hline
96420-01-04-03&$\alpha$&$\rho/\nu$\\
96420-01-05-00&$\nu$&$\rho/\nu$\\
96420-01-06-00&$\rho$&$\rho/\nu$\\
96420-01-07-01&$\rho$&$\mu$\\
96420-01-08-03&$\beta/\lambda$&$\lambda$\\
96420-01-09-06&$\mu$&$\lambda$\\
\hline
\hline

\end{tabular}
\end{table}

\par We acknowledge differences between the classifications assigned by this paper and by \citet{Altamirano_IGR_FH}.  We ascribe these differences to the different approaches we have used to construct our classes.  In particular while we have constructed an independent set of variability classes for IGR J17091 which we have then compared to the \citeauthor{Belloni_GRS_MI} classes for GRS 1915, \citeauthor{Altamirano_IGR_FH} applied the \citeauthor{Belloni_GRS_MI} classes for GRS 1915 directly to IGR J17091.
\par In general, the variability classes we find to be present in IGR J17091 are broadly the same as those noted by \citet{Altamirano_IGR_FH}.  We do not associate any class with Class $\alpha$ in GRS 1915, but we find examples of all of the other variability classes posited by \citeauthor{Altamirano_IGR_FH} to exist in IGR J17091.
\par \citealp{Altamirano_IGR_FH} noted the presence of an anticlockwise loop in the HID of `heartbeat'-like observations of IGR J17091, opposed to the clockwise loop seen in HID of $\rho$-class observations of GRS 1915.  This is consistent with our finding that hysteretic loops in classes III and IV also tend to execute in an anticlockwise direction.  However, we additionally find that hysteretic loops in classes V, VI, VII and VIII tend to execute in a clockwise direction.  This is also different from GRS 1915, in which the loop is executed in the same direction in all classes.  We also additionally report that clockwise loops tend to be more complex than anticlockwise loops seen in IGR J17091, with many showing a multi-lobed structure not seen in GRS 1915.  This apparent inconsistency between the objects strengthens the suggestion in \citealp{Altamirano_IGR_FH} that the heartbeat-like classes in GRS 1915 and IGR J17091 may be generated by physically different mechanisms.

\subsection{New Constraints on Accretion Rate, Mass \& Distance}
\label{sec:newmass}

\par The constraints that \citealp{Altamirano_IGR_FH} placed on the mass and distance of IGR J17091 assumed that the object emitted at its Eddington luminosity at the peak of the 2011--2013 outburst.  They report a peak 2--50\,keV flux of $4\times10^{-9}$\ergf\ during flares in `heartbeat'-like lightcurves during this time.  The correction factor $C_{Bol,Peak}$ to convert 2--50\,keV flux to bolometric flux is not well constrained, but \citealp{Altamirano_IGR_FH} suggest an order-of-magnitude estimate of $\lesssim3$, corresponding to a peak bolometric flux of $\lesssim1.2\times10^{-8}$\ergf .
\par \citealp{Maccarone_2pct} performed a study of the soft to hard transitions in 10 LMXBs.  They found that all but one perform this transition at a luminosity consistent with between 1\% and 4\% of the Eddington limit.  We use \textit{Swift} observation 00031921058 taken on MJD 55965 to create a spectrum of IGR J17091 during the approximate time of its transition from a soft to a hard state \citep{Drave_Return}.  We fit this spectrum above 2\,keV with a power-law, and extrapolate to find a 2--50\,keV flux of $8.56\times10^{-10}$\ergf .  Assuming that the transition bolometric correction factor $C_{Bol,Tran}$ is also $\lesssim3$, this corresponds to a bolometric flux of $\lesssim2.5\times10^{-9}$\ergf .
\par
\begin{figure}
    \includegraphics[width=\columnwidth, trim = 0mm 0mm 0mm 0mm]{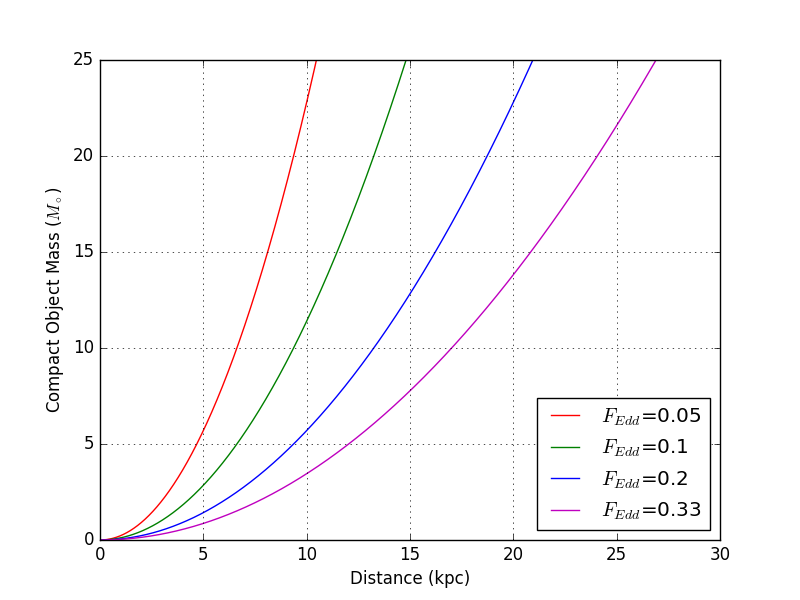}
    \captionsetup{singlelinecheck=off}
    \caption{Mass of the compact object in IGR J17091-3624 plotted against its distance, for values of peak Eddington fractions of $F_{Edd}=$0.05, 0.1, 0.2 and 0.33.}
   \label{fig:IGRMass}
\end{figure}

 By comparing this with the results of \citealp{Maccarone_2pct} and \citealp{Altamirano_IGR_FH}, we find that IGR J17091-3624 was likely emitting at no more than $\sim5$--20\% of its Eddington Limit at its peak.  This number becomes $\sim6\mbox{--}25$\% if we instead use $C_{Bol,Tran}=2.4$, or $\sim8\mbox{--}33$\% if $C_{Bol,Tran}=1.8$.  With this new range of values, we are able to re-derive the compact object mass as the function of the distance (Figure \ref{fig:IGRMass}).  We find that for a black hole mass of $\sim10$\msun , as suggested by \citealp{Iyer_Bayes}, IGR J17091 is within the galaxy at a distance of 6--17\,kpc.  This is consistent with the estimated distance of $\sim11\mbox{--}17$\,kpc estimated by \citealp{Rodriguez_D} for a compact object mass of 10\msun.

\subsection{Implications for Models of `Heartbeat' Variability}

\par We have found that hysteretic HID loops can execute in both directions in IGR J17091 (e.g. Section \ref{sec:Alta}), as well as found a revised estimate that IGR J17091 accretes at $\lesssim20$\% Eddington (Section \ref{sec:newmass}).  Both of these findings have implications for physical models of GRS 1915-like variability in this source.
\par Firstly, we find that Eddington-limited accretion is neither necessary nor sufficient for GRS 1915-like variability.  The discovery of GRS 1915-like variability in the sub-Eddington Rapid Burster \citep{Bagnoli_RB,Bagnoli_PopStudy} provided the first evidence that Eddington-limited accretion may not be a driving factor in this type of variability.  We strengthen this case by finding that IGR J17091-3624 is also likely sub-Eddington.  As such, we further rule out any scenario in which Eddington-limited accretion is required for GRS 1915-like variability in black hole LMXBs specifically.
\par Secondly, by using the direction of hysteretic HID loops, we find that hard photon lag in `heartbeat'-like classes of IGR J17091 can be either positive or negative.  This could mean that we must rule out the causal connection between soft and hard emission being common to all classes.
\par In either case, we find that scenarios that require high global accretion rates or predict a consistent hard photon lag (e.g. \citealp{Neilsen_GRSModel,Janiuk_Lag}), are not able to explain GRS 1915-like variability in IGR J17091 unless they also feature geometric obscuration in a subset of variability classes.  We note that simulations by \citealp{Nayakshin_GRSModel} require an Eddington fraction of $\gtrsim0.26$ before GRS 1915-like variability, a value which falls in the range $\sim0.05\mbox{--}0.33$ that we find for the peak Eddington fraction of IGR J17091.
\par In addition to being near its Eddington limit GRS 1915 also has the largest orbit of any known LMXB (e.g. \citealp{McClintock_BHBs}).  \citealp{Sadowski_MagField} have also shown that thin, radiation dominated regions of disks in LMXBs require a large-scale threaded magnetic field to be stable, and the field strength required to stabilise such a disk in GRS 1915 is higher than for any other LMXB they studied.  We suggest that one of these parameters is more likely to be the criterion for GRS 1915-like variability.  If better constraints can be placed on the disk size and minimum stabilising field strength in IGR J17091, it will become clear whether either of these parameters can be the unifying factor behind LMXBs that display GRS 1915-like variability.

\section{Conclusions}

\par We have constructed the first model-independent set of variability classes for the entire portion of the 2011--2013 outburst of IGR J17091 that was observed with \rxte .  We find that the data are well-described by a set of 9 classes;  7 of these appear to have direct counterparts in GRS 1915, while two are, so far, unique to IGR J17091.  We find that variability in IGR J17091 is generally faster than in the corresponding classes of GRS 1915, and that patterns of quasi-periodic flares and dips form the basis of most variability in both objects.  Despite this, we find evidence that `heartbeat'-like variability in both objects may be generated by different physical processes.  In particular, while hard photons always lag soft in GRS 1915, we find evidence that hard photons can lag or precede soft photons in IGR J17091 depending on the variability class.
\par We also report on the long-term evolution of the 2011--2013 outburst of IGR J17091, in particular noting the presence of 3 re-flares during the later part of the outburst.  Using an empirical relation between hard-soft transition luminosity and Eddington luminosity \citep{Maccarone_2pct}, we estimate that IGR J17091 was likely accreting at no greater than $\sim33$\% of its Eddington limit at peak luminosity.
\par We use these result to conclude that any model of GRS 1915-like variability which requires a near-Eddington global accretion rate is insufficient to explain the variability we see in IGR J17091.  As such we suggest that an extreme value of some different parameter, such as disk size or minimum stabilising large-scale magnetic field, may be the unifying factor behind all objects which display GRS 1915-like variability.  This would explain why sub-Eddington sources such as IGR J17091 and the Rapid Burster do display GRS 1915-like variability, while other Eddington-limited sources such as GX 17+2 and V404 Cyg do not.

\section*{Acknowledgements}

\par J.C. and C.B. thank the Science \& Technology Facilities Council and the Royal Astronomical Society for their financial support.  D.A. thanks the Royal Society, and the International Space Science Institute (\textit{ISSI}) for its support during the `The extreme physics of Eddington and super Eddington accretion onto Black Hole' meetings (\url{http://www.issibern.ch/teams/eddingtonaccretion/}).  T.B. also thanks the \textit{ISSI} for their support.  R.W. is supported by a NWO Top grant, Module 1.  M. Pereyra gratefully acknowledges the Committee on Space Research (COSPAR) Capacity Building and Schlumberger Foundation Faculty for the Future Fellowship Programmes for their financial support.  M.Pahari acknowledges the support of the UGC/UKIERI thematic partnership grant UGC 2014-15/02.
\par The authors acknowledge the use of AstroPy\citep{Astropy}, APLpy \citep{Robitaille_APLpy}, NumPy \citep{NumPy} and MatPlotLib \citep{Hunter_MatPlotLib} libraries for Python.  We also acknowledge the use of public data from the \rxte\ \citep{Bradt_RXTE} archive, as well as \texttt{FTOOLS} \citep{Blackburn_FTools} for data manipulation.
\par This work made use of data supplied by the UK Swift Science Data Centre at the University of Leicester.





\bibliographystyle{mnras}
\bibliography{refs}




\appendix

\section{Flare-Finding Algorithm}
\label{app:Flares}

\par
\begin{figure}
    \includegraphics[width=\columnwidth, trim = 0mm 30mm 0mm 28mm]{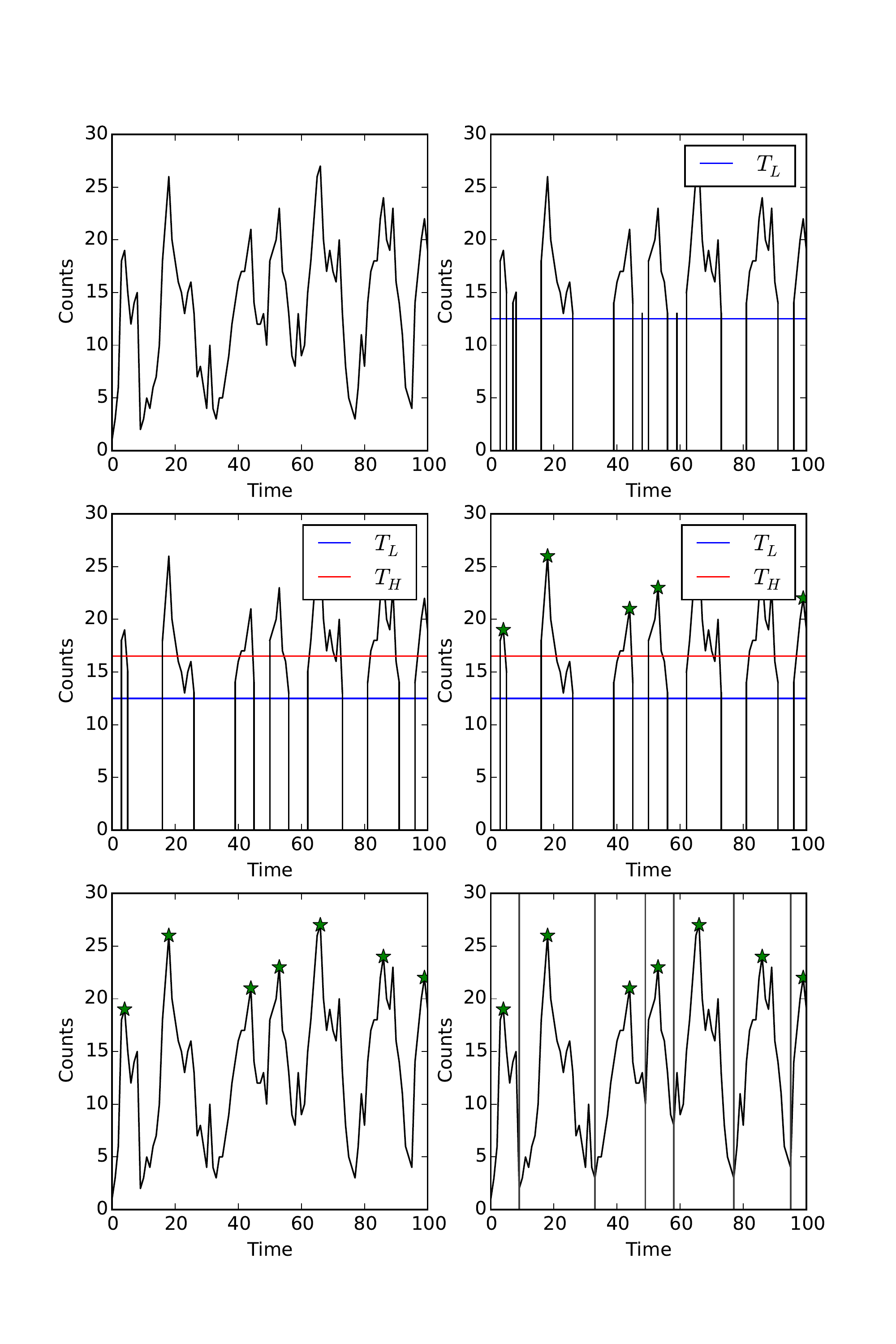}
    \captionsetup{singlelinecheck=off}
    \caption{From top-left: (i) An untouched data-set.  (ii) The dataset with all $y<T_L$ removed.  (iii) The dataset with all contiguous nonzero regions with $\max(y)<T_H$ removed.  (iv) The peak x-values $x_m$.  (v) The restored dataset with the peak x-values $x_m$ highlighted.  (vi) The boundaries between adjacent peaks.}
   \label{fig:BurstAlg}
\end{figure}

 The algorithm used to find flares is performed as such (see also Figure \ref{fig:BurstAlg}):

\begin{enumerate}
  \item Choose some threshold values $T_L$ and $T_H$.  Set the value of all datapoints below $T_L$ to zero.
  \item Retrieve the x-co-ordinate of the highest value remaining in the dataset.  Call this value $x_m$ and store it in a list.
  \item Set the value of point at $x_m$ to zero.
  \item Scan forwards from $x_m$.  If the selected point has a nonzero value, set it to zero and move to the next point.  If the selected point has a zero value, move to step (v).
  \item Scan backwards from $x_m$.  If the selected point has a nonzero value, set it to zero and move to the previous point.  If the selected point has a zero value, move to step (vi).
  \item Retrieve the y-co-ordinate of the highest value remaining in the dataset.  Call this $y_m$.
  \item If $y_m>T_H$, repeat steps (ii)--(vi).  If $y_m<T_H$, proceed to step (viii).
  \item Restore the original dataset.
  \item Retrieve the list of $x_m$ values found in step (ii).  Sort them in order of size.
  \item For each pair of adjacent $x_m$ values, find the x-coordinate of the datapoint between them with the lowest y-value.  Call these values $x_c$.
  \item This list of $x_c$ can now be used to demarcate the border between peaks.
\end{enumerate}

The values $T_L$ and $T_H$ can also be procedurally generated for a given piece of data:

\begin{enumerate}
  \item Select a small section of the dataset or a similar dataset (containing $\sim20$ peaks by eye) and note the location $x_e$ of all peaks found by eye.
  \item Let $P_L$ and $P_H$ be two arbitrary values in the range $[0,100]$.
  \item Let $T_L$ ($T_H$) be the $P_L$th ($P_H$th) percentile of the y-values of the subsection of dataset.
  \item Run the flare-finding algorithm up to step (ix).  Save the list of $x_m$.
  \item Split the dataset into bins on the x-axis such as the bin width $b\ll p$, where $p$ is the rough x-axis separation between peaks.
  \item For each bin, note if you found any value in $x_m$ falls in the bin and note if any value of $x_e$ falls in the bin.
  \item Using each bin as a trial, compute the Heidke Skill Score \citep{Heidke_SKSC} of the algorithm with the method of finding peaks by eye:
  \begin{equation}HSS = \frac{2(AD-BC)}{(A+B)(B+D)+(A+C)(C+D)}
  \label{eq:HSS}
  \end{equation}
  Where $A$ is the number of bins that contain both $x_e$ and $x_m$, $B$ ($C$) is the number of bins that contain only $x_m$ ($x_e$) and $D$ is the number of bins which contain neither \citep{Kok_YesNo}.
  \item Repeat steps (iii)--(vii) for all values of $P_H>P_L$ for $P_L$ and $P_H$ in $[1,100]$.  Use a sensible value for the resolution of $P_L$ and $P_H$.  Save the HSS for each pair of values
  \item Locate the maximum value of HSS, and note the $P_L$ and $P_H$ values used to generate it.  Use these values to generate your final $T_L$ and $T_H$ values.
\end{enumerate}

\begin{figure}
    \includegraphics[width=\columnwidth, trim = 0mm 10mm 0mm 10mm]{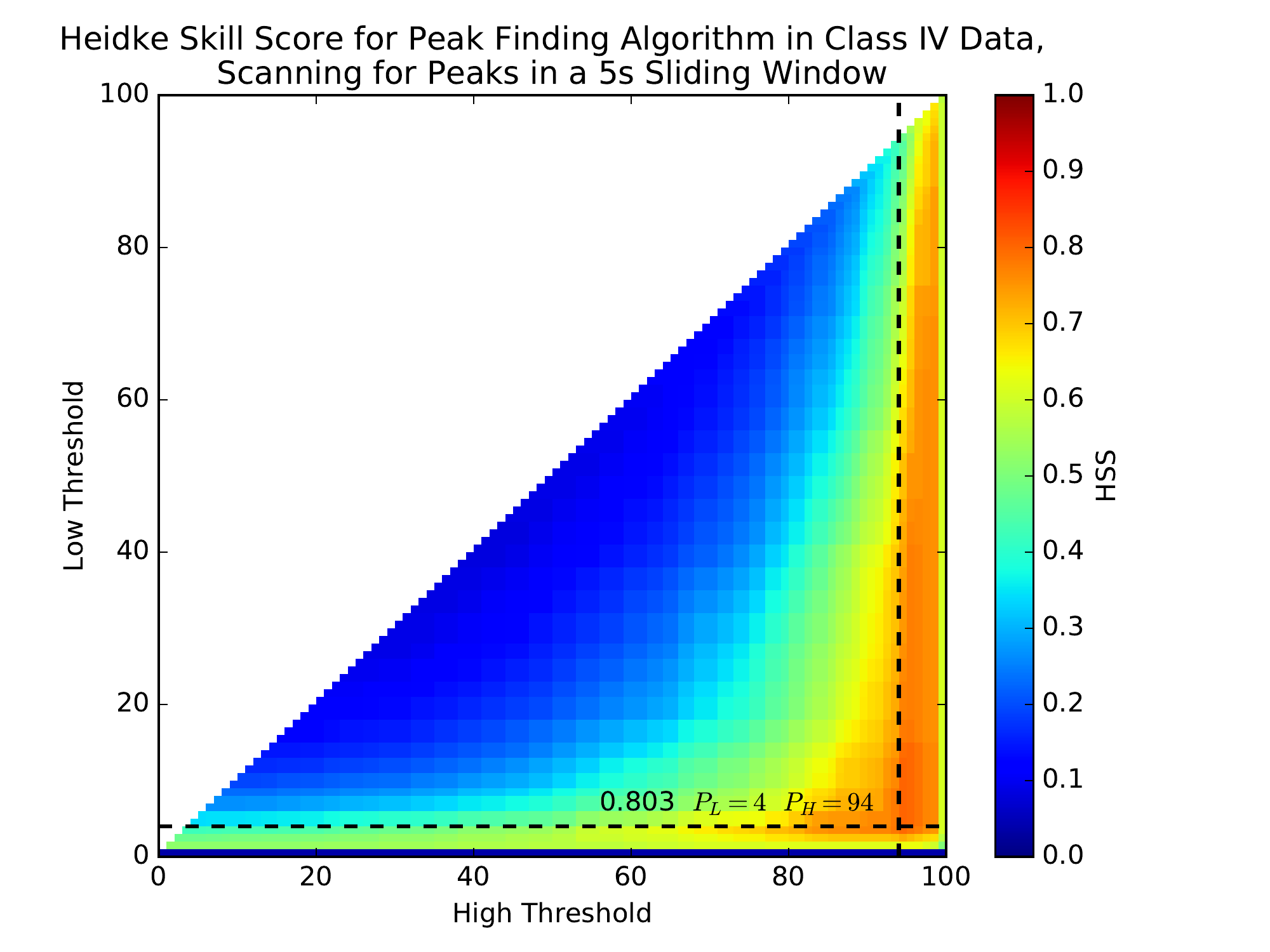}
    \captionsetup{singlelinecheck=off}
    \caption{The Heidke Skill score of a Class IV observation of IGR J17091-3624 for a selection of different values $P_L$ and $P_H$.}
   \label{fig:Heidke}
\end{figure}

We show an example of Heidke skill score grid for this algorithm, applied to a Class IV observation, in Figure \ref{fig:Heidke}.

\section{Model-Independent Classification of each Observation of IGR J17091-3624}
\label{app:Obsids}

\par Observation IDs, and orbit IDs, for every observation and observation segment that was used in our analysis are presented in Table~\ref{tab:obsids}.  Note that not all of every observation was used; in many cases, large spikes caused by \pca\ PCUs switching off or on rendered $\sim100$\,s unusable.  As these often occurred very close to the beginning or end of an observation segment, small sections of data before or after these spikes was also sometimes discarded.  Every observation segment is presented along with the variability class assigned to it by this study.

\begin{table*}
\caption{Here is listed the Observation IDs for every $RXTE$ observation that was used in this analysis, along with the variability class which has been assigned to it.  \textit{Orb.} is the orbit ID (starting at 0) of each observation segment, \textit{Exp.} is the exposure time in seconds and \textbf{X} is the prefix 96420-01.  This table is continued overleaf in Table \ref{tab:obsids2}.}
\label{tab:obsids}
\begin{tabular}{llllrllllrllllr}
\hline
\hline
MJD&OBSID&\textit{Orb.}&Class&\textit{Exp.}&MJD&OBSID&\textit{Orb.}&Class&\textit{Exp.}&MJD&OBSID&\textit{Orb.}&Class&\textit{Exp.}\\
\hline
55622&\textbf{X}-01-00&0&I&1840&55676&\textbf{X}-09-06&0&VI&3540&55741&\textbf{X}-18-05&0&VII&782\\
55622&\textbf{X}-01-000&0&I&3480&55677&\textbf{X}-09-01&0&V&1676&55743&\textbf{X}-19-00&0&VII&1412\\
55622&\textbf{X}-01-000&1&I&1656&55678&\textbf{X}-09-04&0&V&2090&55744&\textbf{X}-19-01&0&VIII&1938\\
55622&\textbf{X}-01-000&2&I&3384&55679&\textbf{X}-09-02&0&V&2306&55745&\textbf{X}-19-02&0&VII&2172\\
55622&\textbf{X}-01-000&3&I&3400&55680&\textbf{X}-10-02&0&V&952&55747&\textbf{X}-19-03&0&VIII&1691\\
55622&\textbf{X}-01-000&4&I&3384&55681&\textbf{X}-10-00&0&V&3725&55748&\textbf{X}-19-04&0&VI&1283\\
55623&\textbf{X}-01-01&0&I&1240&55682&\textbf{X}-10-03&0&V&1157&55749&\textbf{X}-19-05&0&VIII&1417\\
55623&\textbf{X}-01-01&1&I&752&55684&\textbf{X}-10-01&0&III&1504&55751&\textbf{X}-20-05&0&VI&1726\\
55623&\textbf{X}-01-01&2&I&992&55686&\textbf{X}-10-04&0&III&1127&55752&\textbf{X}-20-01&0&VIII&1079\\
55623&\textbf{X}-01-01&3&I&1184&55686&\textbf{X}-10-05&0&II&2179&55753&\textbf{X}-20-02&0&VIII&1433\\
55623&\textbf{X}-01-01&4&I&1056&55687&\textbf{X}-11-00&0&II&3537&55754&\textbf{X}-20-03&0&VII&1122\\
55623&\textbf{X}-01-010&0&I&2080&55688&\textbf{X}-11-01&0&II&1153&55756&\textbf{X}-20-04&0&VIII&1486\\
55623&\textbf{X}-01-010&1&I&1832&55690&\textbf{X}-11-02&0&II&1408&55757&\textbf{X}-21-00&0&VIII&3372\\
55623&\textbf{X}-01-010&2&I&1648&55691&\textbf{X}-11-03&0&II&886&55758&\textbf{X}-21-01&0&VIII&3383\\
55623&\textbf{X}-01-010&4&I&1424&55692&\textbf{X}-11-04&0&II&3566&55759&\textbf{X}-21-02&0&VI&1938\\
55623&\textbf{X}-01-010&5&I&400&55693&\textbf{X}-11-05&0&II&1817&55761&\textbf{X}-21-04&0&VII&1497\\
55623&\textbf{X}-01-02&0&I&3056&55694&\textbf{X}-12-00&0&II&2761&55762&\textbf{X}-21-05&0&VII&1548\\
55623&\textbf{X}-01-02&1&I&2792&55695&\textbf{X}-12-01&0&II&1374&55763&\textbf{X}-21-06&0&VII&2202\\
55623&\textbf{X}-01-02&2&I&2432&55695&\textbf{X}-12-02&0&II&2041&55764&\textbf{X}-22-00&0&VII&1682\\
55623&\textbf{X}-01-020&0&I&3456&55696&\textbf{X}-12-03&0&II&1456&55765&\textbf{X}-22-01&0&VII&1221\\
55623&\textbf{X}-01-020&1&I&3464&55698&\textbf{X}-12-04&0&II&1916&55766&\textbf{X}-22-02&0&V&720\\
55623&\textbf{X}-01-020&2&I&3512&55698&\textbf{X}-12-05&0&II&3139&55767&\textbf{X}-22-03&0&V&1801\\
55623&\textbf{X}-01-020&3&I&3520&55700&\textbf{X}-12-06&0&II&1189&55768&\textbf{X}-22-04&0&VIII&1983\\
55623&\textbf{X}-01-020&4&I&3512&55701&\textbf{X}-13-00&0&II&1214&55769&\textbf{X}-22-05&0&VIII&999\\
55623&\textbf{X}-01-020&5&I&464&55702&\textbf{X}-13-01&0&II&980&55770&\textbf{X}-22-06&0&VIII&667\\
55624&\textbf{X}-02-00&0&I&1758&55704&\textbf{X}-13-02&0&II&732&55771&\textbf{X}-23-00&0&VIII&2075\\
55626&\textbf{X}-02-01&0&I&1380&55705&\textbf{X}-13-03&0&III&1217&55772&\textbf{X}-23-01&0&VII&3385\\
55628&\textbf{X}-02-02&0&I&3305&55706&\textbf{X}-13-04&0&III&1161&55773&\textbf{X}-23-02&0&VII&2218\\
55630&\textbf{X}-02-03&0&I&1876&55707&\textbf{X}-13-05&0&IV&2763&55774&\textbf{X}-23-03&0&V&1811\\
55632&\textbf{X}-03-00&0&I&1712&55708&\textbf{X}-14-00&0&IV&1188&55775&\textbf{X}-23-04&0&V&3356\\
55634&\textbf{X}-03-01&0&III&3590&55709&\textbf{X}-14-01&0&IV&3342&55776&\textbf{X}-23-05&0&V&2603\\
55639&\textbf{X}-04-00&0&IV&3099&55710&\textbf{X}-14-02&0&IV&1094&55777&\textbf{X}-23-06&0&IV&912\\
55642&\textbf{X}-04-02&0&IV&2972&55712&\textbf{X}-14-03&0&IV&1404&55777&\textbf{X}-23-06&1&IV&1544\\
55643&\textbf{X}-04-01&0&III&1190&55713&\textbf{X}-14-04&0&V&871&55778&\textbf{X}-24-00&0&IV&1309\\
55644&\textbf{X}-04-03&0&III&2903&55714&\textbf{X}-14-05&0&V&1311&55779&\textbf{X}-24-01&0&IV&3599\\
55645&\textbf{X}-05-02&0&I&3578&55715&\textbf{X}-15-00&0&IV&1241&55779&\textbf{X}-24-02&0&IV&2013\\
55647&\textbf{X}-05-00&0&IV&2872&55716&\textbf{X}-15-01&0&IV&1262&55782&\textbf{X}-24-03&0&V&1761\\
55647&\textbf{X}-05-000&0&IV&3472&55717&\textbf{X}-15-02&0&III&1557&55782&\textbf{X}-24-04&0&V&1725\\
55647&\textbf{X}-05-000&1&IV&3520&55718&\textbf{X}-15-03&0&III&1334&55784&\textbf{X}-24-05&0&V&3144\\
55647&\textbf{X}-05-000&2&IV&3512&55720&\textbf{X}-15-04&0&IV&1486&55784&\textbf{X}-24-06&0&V&2591\\
55647&\textbf{X}-05-000&3&IV&3520&55721&\textbf{X}-15-05&0&IV&1500&55785&\textbf{X}-25-00&0&V&2366\\
55647&\textbf{X}-05-000&4&IV&3512&55722&\textbf{X}-16-00&0&IV&900&55786&\textbf{X}-25-01&0&V&1804\\
55647&\textbf{X}-05-000&5&IV&648&55723&\textbf{X}-16-01&0&III&1004&55787&\textbf{X}-25-02&0&V&1951\\
55649&\textbf{X}-05-03&0&IV&2409&55724&\textbf{X}-16-02&0&II&1923&55788&\textbf{X}-25-03&0&V&1619\\
55650&\textbf{X}-05-01&0&IV&1473&55725&\textbf{X}-16-03&0&II&1919&55789&\textbf{X}-25-04&0&V&2601\\
55651&\textbf{X}-05-04&0&IV&2954&55726&\textbf{X}-16-04&0&III&1935&55790&\textbf{X}-25-05&0&V&1473\\
55653&\textbf{X}-06-00&0&IV&2723&55727&\textbf{X}-16-05&0&II&730&55791&\textbf{X}-25-06&0&V&922\\
55654&\textbf{X}-06-01&0&IV&3388&55728&\textbf{X}-16-06&0&II&1953&55792&\textbf{X}-26-00&0&V&2336\\
55656&\textbf{X}-06-02&0&IV&2908&55729&\textbf{X}-17-00&0&II&2735&55794&\textbf{X}-26-01&0&V&1385\\
55657&\textbf{X}-06-03&0&V&1842&55730&\textbf{X}-17-01&0&II&3556&55795&\textbf{X}-26-02&0&VIII&1458\\
55661&\textbf{X}-07-00&0&V&1754&55731&\textbf{X}-17-02&0&II&3605&55796&\textbf{X}-26-03&0&VI&1325\\
55662&\textbf{X}-07-01&0&V&3365&55732&\textbf{X}-17-03&0&II&1647&55798&\textbf{X}-26-04&0&VI&2075\\
55663&\textbf{X}-07-02&0&V&3373&55733&\textbf{X}-17-04&0&II&1459&55799&\textbf{X}-27-00&0&VI&1396\\
55666&\textbf{X}-08-00&0&V&3338&55734&\textbf{X}-17-05&0&III&1736&55800&\textbf{X}-27-01&0&VI&2684\\
55669&\textbf{X}-08-01&0&V&3368&55735&\textbf{X}-17-06&0&III&3653&55801&\textbf{X}-27-02&0&VI&1016\\
55670&\textbf{X}-08-03&0&VI&2489&55736&\textbf{X}-18-00&0&III&2317&55802&\textbf{X}-27-03&0&VI&1179\\
55671&\textbf{X}-08-02&0&VI&2609&55737&\textbf{X}-18-01&0&IV&1387&55803&\textbf{X}-27-04&0&VI&1304\\
55673&\textbf{X}-09-03&0&VI&1011&55738&\textbf{X}-18-02&0&V&1291&55805&\textbf{X}-27-05&0&VI&1663\\
55674&\textbf{X}-09-00&0&VI&1386&55739&\textbf{X}-18-03&0&V&2178&55806&\textbf{X}-28-00&0&VI&1456\\
55675&\textbf{X}-09-05&0&IX&1148&55740&\textbf{X}-18-04&0&V&1478&55808&\textbf{X}-28-01&0&VIII&577\\
\hline
\hline
\end{tabular}

\end{table*}

\begin{table*}
\caption{A continuation of Table \ref{tab:obsids}.  \textit{Orb.} is the orbit ID (starting at 0) of each observation segment, \textit{Exp.} is the exposure time in seconds and \textbf{X} is the prefix 96420-01.}
\label{tab:obsids2}
\begin{tabular}{llllrllllrllllr}
\hline
\hline
MJD&OBSID&\textit{Orb.}&Class&\textit{Exp.}&MJD&OBSID&\textit{Orb.}&Class&\textit{Exp.}&MJD&OBSID&\textit{Orb.}&Class&\textit{Exp.}\\
\hline
55810&\textbf{X}-28-02&0&VI&1251&55836&\textbf{X}-32-02&0&IX&1591&55857&\textbf{X}-35-01&0&IX&1912\\
55811&\textbf{X}-28-03&0&VI&2000&55837&\textbf{X}-32-03&0&IX&2155&55859&\textbf{X}-35-02&0&IX&200\\
55813&\textbf{X}-29-00&0&VIII&1309&55838&\textbf{X}-32-04&0&IX&2641&55859&\textbf{X}-35-02&1&IX&1296\\
55819&\textbf{X}-29-04&0&VIII&1686&55838&\textbf{X}-32-05&0&IX&2077&55860&\textbf{X}-35-03&0&IX&1372\\
55820&\textbf{X}-30-00&0&VI&1488&55840&\textbf{X}-32-06&0&IX&3392&55861&\textbf{X}-35-04&0&IX&836\\
55821&\textbf{X}-30-01&0&VI&1503&55840&\textbf{X}-32-06&1&IX&3512&55862&\textbf{X}-36-00&0&IX&1145\\
55822&\textbf{X}-30-02&0&VI&1417&55840&\textbf{X}-32-06&2&IX&3934&55863&\textbf{X}-36-01&0&IX&1322\\
55823&\textbf{X}-30-03&0&VI&1290&55840&\textbf{X}-32-06&3&IX&3880&55865&\textbf{X}-36-03&0&IX&1485\\
55824&\textbf{X}-30-04&0&VI&1489&55840&\textbf{X}-32-06&4&IX&1896&55866&\textbf{X}-36-04&0&IX&1795\\
55825&\textbf{X}-30-05&0&VI&2581&55841&\textbf{X}-33-00&0&IX&1188&55867&\textbf{X}-36-05&0&IX&1732\\
55826&\textbf{X}-30-06&0&VI&2747&55842&\textbf{X}-33-01&0&IX&855&55868&\textbf{X}-36-06&0&IX&1657\\
55827&\textbf{X}-31-00&0&VI&1559&55843&\textbf{X}-33-02&0&IX&1156&55871&\textbf{X}-37-00&0&IX&815\\
55828&\textbf{X}-31-01&0&VI&2954&55845&\textbf{X}-33-04&0&IX&1713&55871&\textbf{X}-37-02&0&IX&1460\\
55829&\textbf{X}-31-02&0&IX&3005&55846&\textbf{X}-33-05&0&IX&934&55872&\textbf{X}-37-03&0&IX&1683\\
55830&\textbf{X}-31-03&0&IX&1472&55847&\textbf{X}-33-06&0&IX&717&55873&\textbf{X}-37-04&0&IX&1402\\
55830&\textbf{X}-31-03&1&IX&288&55848&\textbf{X}-34-00&0&IX&1159&55874&\textbf{X}-37-05G&0&IX&1536\\
55831&\textbf{X}-31-04&0&IX&1586&55849&\textbf{X}-34-01&0&IX&973&55875&\textbf{X}-37-06&0&IX&1536\\
55832&\textbf{X}-31-05&0&VI&3812&55851&\textbf{X}-34-02&0&IX&2261&55876&\textbf{X}-38-00&0&IX&1497\\
55833&\textbf{X}-31-06&0&IX&3675&55852&\textbf{X}-34-03&0&IX&1092&55877&\textbf{X}-38-01&0&IX&1134\\
55834&\textbf{X}-32-00&0&IX&1217&55853&\textbf{X}-34-04&0&IX&741&55878&\textbf{X}-38-02&0&IX&1289\\
55835&\textbf{X}-32-01&0&IX&1445&55856&\textbf{X}-35-00&0&IX&797&55879&\textbf{X}-38-03&0&IX&1433\\
\hline
\hline
\end{tabular}
\end{table*}


\bsp	
\label{lastpage}
\end{document}